\documentclass[sn-mathphys,iicol]{sn-jnl}

\usepackage[absolute]{textpos}
\textblockorigin{10mm}{10mm}

\jyear{2022}%

\theoremstyle{thmstyleone}%
\theoremstyle{thmstyletwo}%

\theoremstyle{thmstylethree}%

\begin{document}

\title[PDE-based Lossless Compression of Diffusion MR Images]{Combining Image Space and q-Space PDEs for Lossless Compression of Diffusion MR Images}


\author{\fnm{Ikram} \sur{Jumakulyyev}}\email{s6ikjuma@uni-bonn.de}
\author*{\fnm{Thomas} \sur{Schultz}}\email{schultz@cs.uni-bonn.de}

\affil{\orgdiv{University of Bonn}, \orgname{B-IT and Department of Computer Science II}, \orgaddress{\street{Friedrich-Hirzebruch-Allee 5}, \city{Bonn}, \postcode{53115}, \state{} \country{Germany}}}

\abstract{Diffusion MRI is a modern neuroimaging modality with a unique ability to acquire microstructural information by measuring water self-diffusion at the voxel level. However, it generates huge amounts of data, resulting from a large number of repeated 3D scans. Each volume samples a location in q-space, indicating the direction and strength of a diffusion sensitizing gradient during the measurement. This captures detailed information about the self-diffusion, and the tissue microstructure that restricts it. Lossless compression with GZIP is widely used to reduce the memory requirements. We introduce a novel lossless codec for diffusion MRI data. It reduces file sizes by more than 30\% compared to GZIP, and also beats lossless codecs from the JPEG family. Our codec builds on recent work on lossless PDE-based compression of 3D medical images, but additionally exploits smoothness in q-space. We demonstrate that, compared to using only image space PDEs, q-space PDEs further improve compression rates. Moreover, implementing them with Finite Element Methods and a custom acceleration significantly reduces computational expense. Finally, we show that our codec clearly benefits from integrating subject motion correction, and slightly from optimizing the order in which the 3D volumes are coded.}




\maketitle

\begin{textblock}{5}(0.25,0)
This version of the article has been accepted for publication, after peer review, but is not the Version of Record and does not reflect post-acceptance improvements, or any corrections. The Version of Record is available online at: \url{https://doi.org/10.1007/s10851-023-01144-z}
\end{textblock}
\section{Introduction}
\label{sec:intro}
With the development of new medical imaging techniques, and constant refinement of existing ones, the associated storage requirements have been reported to grow exponentially each year \cite{dinov2016volume}. This explains why medical image compression is an active area of research.

Our work belongs to the family of compression algorithms that are based on Partial Differential Equations (PDEs). The general idea behind this approach is to store a sparse subset of the image information, and to reconstruct the remaining image via PDE-based inpainting \cite{galic2008image}.

PDE-based compression has a long tradition for the lossy compression of natural images \cite{galic2008image,schmaltz2014understanding} and videos \cite{peter2015beyond,kostler2007pde,andris2016proof}. The benefit of PDE-based approaches relative to transform-based codecs like JPEG \cite{10.5555/573326} and JPEG2000 \cite{Taubman:2002} has often been most pronounced at high compression rates \cite{schmaltz2014understanding}. Even though this strategy for lossy compression has also been transferred to three-dimensional images \cite{peter2013three}, in medical imaging, lossless compression is often preferred to ensure that all diagnostically relevant details are preserved. In some cases, it is even legally forbidden to apply lossy compression for medical image archival \cite{kil2006lossless,miaou2009lossless}.

We recently introduced a PDE-based codec for 3D medical images that stores the residuals between the original image and an intermediate PDE-based reconstruction to ensure that the final reconstruction is lossless, and we demonstrated that this strategy led to competitive compression rates \cite{jumakulyyev2021lossless}. In our current work, we extend this idea for the specific use case of image datasets from diffusion MRI.

Diffusion MRI (dMRI) \cite{LeBihan:1986,Basser:1994} is a variant of Magnetic Resonance Imaging in which diffusion sensitizing gradients are introduced into the measurement sequence. If the hydrogen nuclei that generate the MR signal undergo a net displacement along the gradient direction during the measurement, the signal is attenuated. Assuming that these displacements result from (self-) diffusion, comparing diffusion-weighted to non-weighted measurements permits computation of an apparent diffusion coefficient.

Taking measurements with different gradient directions captures the directional dependence of the diffusivity. It results from interactions between water and tissue microstructure and therefore carries information about structures that are much smaller than the MR image resolution. Important applications of dMRI include the detection of microstructural changes that are related to aging or disease, and the reconstruction of major white matter tracts, which is referred to as fiber tracking or tractography \cite{Jones:2011Book}.

The large number of repeated measurements in diffusion MRI leads to large
amounts of data. In practice, resulting image datasets are often compressed using GZIP \cite{10.17487/RFC1952}. In our previous work \cite{jumakulyyev2021lossless}, we demonstrated that, compared to this, PDE-based lossless compression can further reduce the memory requirement of individual dMRI volumes by more than 25\%. However, applying our codec to each 3D volume independently does not exploit the fact that measurements for nearby gradient directions are usually similar. Moreover, it is relatively time consuming.

In our current work, we address both of these limitations by combining the previous idea of lossless compression via image-space inpainting with a novel approach of PDE-based inpainting in q-space, which is the space spanned by diffusion sensitizing gradient directions and magnitudes. We find that predictions from linear diffusion in q-space can be made with low computational effort, and are strong enough to further improve compression rates.

The remainder of our work is organized as follows: Section~\ref{sec:related-work} provides the required background and discusses prior work on 4D image compression. Section~\ref{sec:codec} introduces the components of our proposed codec. Section~\ref{sec:results} demonstrates that the resulting compression rates exceed those of several baselines and investigates the effects of specific design choices. Section~\ref{sec:conclusion} concludes with a brief discussion.

\section{Background and Related Work}
\label{sec:related-work}
We will now introduce the main ideas behind diffusion PDE-based image inpainting and compression (Section~\ref{subsec:diffusionPDEs}), clarify the foundations of diffusion MRI and q-space (Section~\ref{subsec:dMRI}), and briefly review the literature on 4D medical image compression (Section~\ref{subsec:4d-medImg-compression}).

\subsection{Diffusion PDE-based Inpainting and Compression}
\label{subsec:diffusionPDEs}
Inspired by their use for modeling physical phenomena, Partial Differential Equations (PDEs) have a long tradition for solving problems in image processing.
In particular, the PDE describing heat diffusion has provided a framework for image smoothing and inpainting
\cite{iijima1962basic,perona1990scale,osher1990feature,weickert1998anisotropic,masnou1998level,bertalmio2000image,chan2001nontexture}.

The heat equation captures the relationship between temporal changes in a temperature $\partial_t u$ and the divergence of its spatial gradient $\nabla u$,
\begin{equation} \label{eq:heatModel}
	{\partial_t u} = \mathrm{div}(D \cdot \nabla u) \;  ,
\end{equation}
where $D$ is the thermal diffusivity of the medium. In a homogeneous and isotropic medium, the diffusivity $D$ is a constant scalar. In a non-homogeneous isotropic medium, $D$ would still be a scalar, but depend on the spatial location. In an anisotropic medium, heat dissipates more rapidly in some directions than in others. In that case, $D$ is a symmetric positive definite matrix that is referred to as a diffusion tensor.

In image processing, the gray value at a certain location is interpreted as a temperature $u$, and Equation~(\ref{eq:heatModel}) is coupled with suitable boundary conditions. For image smoothing,
\begin{equation} \label{eq:smoothingModel}
\begin{split}
	{\partial_t u} &= \mathrm{div}(D \cdot \nabla u), \quad \Omega \times (0,\infty)\ , \\
	{\partial_n u} &= 0, \quad \partial\Omega \times (0,\infty) \ ,
\end{split}
\end{equation}
where $\Omega$ is the image domain, and $n$ is the normal vector to its boundary $\partial\Omega$. The original image $f:\Omega\rightarrow\mathbb{R}$ is used to specify an initial condition $u=f$ at $t=0$. For increasing diffusion time $t$, $u$ will correspond to an increasingly smoothed version of the image.

In image inpainting, values are known at a subset of pixel locations, and unknown values should be filled in. For this, a Dirichlet boundary condition is introduced, which fixes values at a subset $K$ of pixel locations \cite{weickert1998anisotropic,galic2008image}
\begin{equation} \label{eq:inpaintingModel}
\begin{split}
	{\partial_t u} &= \mathrm{div}(D \cdot \nabla u), \quad \Omega \backslash K \times (0,\infty)\ , \\
	{\partial_n u} &= 0, \quad \partial\Omega \times (0,\infty) \ , \\
	u &= f, \quad K\times [0,\infty) \ .
\end{split}
\end{equation}
and a steady-state is computed at which $\partial_t u\approx 0$. The ability of PDEs to reconstruct plausible images even from a very sparse subset of pixels made them useful for image compression \cite{galic2008image,schmaltz2014understanding,peter2015beyond}. 

Different choices of diffusivity $D$ introduce considerable flexibility with respect to shaping the final result. Fixing $D=1$ turns Equation~(\ref{eq:inpaintingModel}) into second-order linear homogenous (LH) diffusion
\begin{equation} \label{eq:linearInpaintingModel}
  {\partial_t u} = L u, \quad \Omega \backslash K \times (0,\infty)
\end{equation}
with $Lu=\Delta u$, where $\Delta$ denotes the Laplace operator, and the steady state satisfies the Laplace equation $\Delta u = 0$. Even though the resulting reconstructions suffer from singularities \cite{galic2008image} and can often be improved by the more complex models discussed below, they have been used to design compression codecs for cartoon-like images \cite{mainberger2011edge}, flow fields \cite{jost2020compressing}, and depth maps \cite{gautier2012efficient,hoffmann2013compression,li2012scalable}. Its simple linear nature and fast convergence to the steady-state also make LH diffusion an attractive choice for real-time video compression \cite{kostler2007pde,andris2016proof}. 

Compared to LH diffusion, decreasing the diffusivity as a function of image gradient magnitude permits a better preservation of salient edges \cite{perona1990scale,charbonnier1997deterministic}. This is referred to as nonlinear diffusion, since the results are no longer linear in the original image $f$.
Rather than just decreasing the overall diffusivity close to edges, modeling $D$ as an anisotropic diffusion tensor permits smoothing along edges, while maintaining or even increasing the contrast perpendicular to them. One widely used model is referred to as Edge-Enhancing Diffusion (EED) \cite{10.1007/978-3-7091-6586-7_13}.

All PDEs that have been discussed up to this point are of second order. Fourth- and higher-order extensions have also been studied, both for smoothing \cite{you2000fourth,lysaker2003noise,didas2009properties,hajiaboli2011anisotropic,zadeh2017multi} and for inpainting \cite{li2013two,10.1007/978-3-030-56215-1_5}. In the simplest case, setting $Lu=-\Delta^2 u$ in Equation~(\ref{eq:linearInpaintingModel}) leads to the biharmonic (BH) equation. In two and three dimensions, it does not suffer from the singularities that are present in the results of LH diffusion \cite{chen2014bi,galic2008image}, while preserving a simple linear nature. For this reason, BH has been considered for the design of compression codecs \cite{peter2016evaluating,chen2014bi,amrani2017diffusion,peter2016turning}. However, it no longer satisfies a min-max principle \cite{didas2009properties} and it increases running time and sensitivity to quantization error.

Our own previous work \cite{10.1007/978-3-030-56215-1_5} proposed an anisotropic fourth-order PDE in which a fourth-order diffusion tensor is constructed from the image gradient in a similar way as in second-order EED. We thus refer to it as Fourth-Order Edge-Enhancing Diffusion (FOEED). It was shown to result in more accurate inpainting results than second-order EED, and higher PDE-based compression rates, in several examples \cite{jumakulyyev2021lossless}.

Our current work is concerned with compressing data from diffusion MRI, which is similar to hyperspectral images in that it contains a large number of values (channels) at each location \cite{amrani2017diffusion}. However, the channels in hyperspectral images have a one-dimensional order, while our channels correspond to positions on a spherical shell in q-space, which we will now introduce.

\subsection{Diffusion MRI}
\label{subsec:dMRI}

The signal in Magnetic Resonance Imaging is generated by the hydrogen atoms within water molecules. Their heat motion is referred to as self-diffusion, since it takes place despite a zero concentration gradient. The extent to which this motion is restricted in a cellular environment correlates with microstructural parameters such as cellular density or integrity. Moreover, in the white matter of the human brain, which contains the tracts that connect different brain regions, self-diffusion can occur more freely in the local direction of those tracts than perpendicular to it \cite{Beaulieu:2002}. Therefore, measuring the apparent diffusion coefficient in different directions provides relevant information about small-scale structures that are below the image resolution of in vivo MRI.

This motivates the use of diffusion MRI. It goes back to the idea of measuring diffusion by introducing a pair of diffusion sensitizing magnetic field gradients into a nuclear magnetic resonance sequence \cite{stejskal1965spin}. Integrating it with spatially resolved Magnetic Resonance Imaging permits diffusion measurements at a voxel level \cite{LeBihan:1986}. Repeating the measurements with differently oriented gradients reveals a biologically relevant directional dependence in various tissue types, including muscle and the white matter of the brain \cite{Pierpaoli:1996a}.

Several key parameters of the diffusion sensitization can be summarized in the gradient wave vector
\begin{equation}
  \label{eq:q-vector}
  \mathbf{q}=\frac{1}{2\pi}\gamma\delta\mathbf{g},
\end{equation}
where $\gamma$ is the gyromagnetic ratio of hydrogen nuclei in water, $\delta$ is the duration of the diffusion sensitizing gradients, and $\mathbf{g}$ corresponds to their direction and strength. The normalized MR echo magnitude $\lvert E(\mathbf{q},\tau)\rvert$ additionally depends on the time $\tau$ between the pair of gradient pulses. It is computed as the ratio between the corresponding diffusion-weighted measurement and an unweighted measurement with $\mathbf{q}=\mathbf{0}$. It is antipodally symmetric, $\lvert E(\mathbf{-q},\tau)\rvert=\lvert E(\mathbf{q},\tau)\rvert$.

The relevance of this q-space formalism derives from a Fourier relationship between $\lvert E(\mathbf{q},\tau)\rvert$ and the ensemble average diffusion propagator $\bar{P}(\mathbf{R},\tau)$, which specifies the probability of a molecular displacement $\mathbf{R}$ within a fixed diffusion time \cite{Callaghan:1988}. An alternative parameterization of the diffusion gradients is in terms of their direction and a factor $b=4\pi^2\|\mathbf{q}\|^2\left(\tau-\delta/3\right)$, which also accounts for the fact that the diffusion weighting increases with the effective diffusion time $\left(\tau-\delta/3\right)$.

\begin{figure}
  \centering
  \includegraphics[height=.175\textheight, width=.492\linewidth]{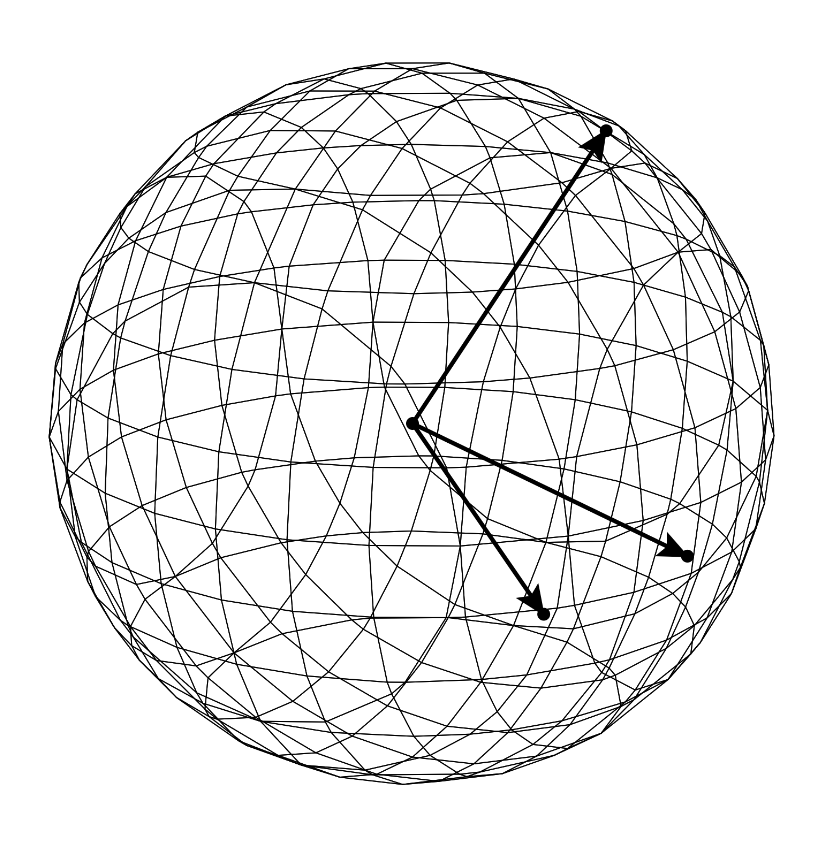}
  \hfill
  \includegraphics[height=.17\textheight, width=.492\linewidth]{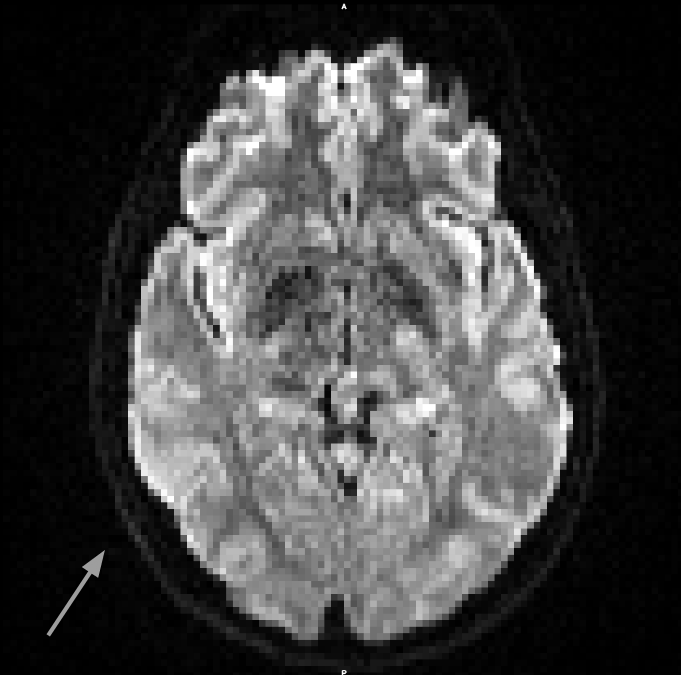}
  \centering
  \includegraphics[height=.17\textheight, width=.492\linewidth]{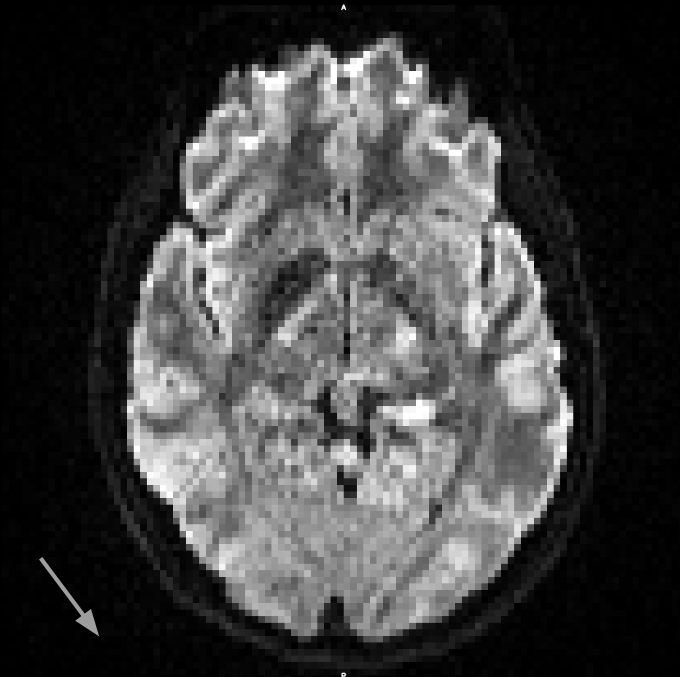}
  \includegraphics[height=.17\textheight, width=.492\linewidth]{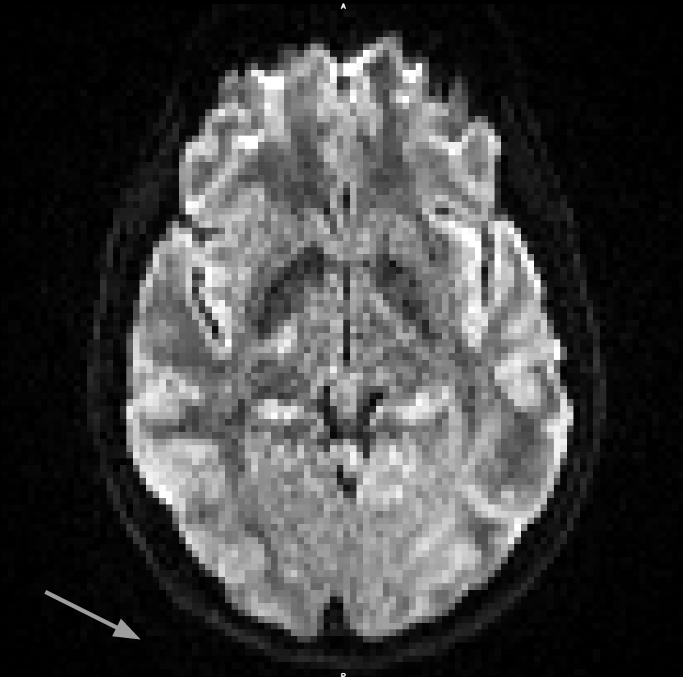}
  \hfill
  \caption{\label{fig:magneticField-gradient-diffusion}%
    Illustration of three diffusion sensitizing gradient directions on a shell in $q$ space, with equal $b=700$ (top left). The three diffusion-weighted images have been measured with different gradient directions, as indicated at the bottom left of each image. Comparing them reveals the directional dependence of the dMRI signal.}
\end{figure}

Due to practical constraints on the overall duration of dMRI measurements, the sampling of q-space is usually limited to one or several reference measurements with $\mathbf{q}=\mathbf{0}$, as well as one or a few shells with constant $\|\mathbf{q}\|$, and thus constant $b$. This is illustrated in Figure~\ref{fig:magneticField-gradient-diffusion}. Such setups focus on the directional dependence of the signal, and typically strive for a uniform distribution of gradient directions on these shells \cite{cheng2017single}. Our codec assumes dMRI data with such a ``shelled'' structure, an assumption that is shared by well-established algorithms in the field \cite{Andersson2016a}.

\subsection{4D Medical Image Compression}
\label{subsec:4d-medImg-compression}

Many medical imaging modalities, including Magnetic Resonance Imaging, Computed Tomography, and ultrasound, can be used to image volumes repeatedly, in order to capture time-dependent phenomena such as organ motion, perfusion, or blood oxygenation. Considerable work has been done on lossless and lossy compression of the resulting 4D (3D plus time) image data. Much of it has borrowed from video coding, and has often involved motion compensation \cite{kassim2005motion,sanchez2008efficient}, which is combined with wavelet transforms \cite{zeng2002four,lalgudi2005compression,kassim2005motion,liu2007four,belhadef2016lossless} or hierarchical vector quantization \cite{nguyen2011efficient} for compression.

Almost all of these works have compared their compression rates to codecs from the JPEG family. We will also compare our codec to JPEG-LS, lossless JPEG, and JPEG2000. Additionally, we compare compression rates against GZIP \cite{10.17487/RFC1952} which, in conjunction with the Neuroimaging Informatics Technology Initiative (NIfTI) file format, is currently most widely used to compress diffusion MRI data in practice. To make this comparison fair, we also use Huffman coding or Deflate within our own codec, as opposed to computationally efficient alternatives that might further improve compression rates \cite{Witten:1987,Duda:2015}.

Even though the volumetric images in diffusion MRI are also taken sequentially, their temporal order is less relevant than the q-space structure that was described above: Measuring with the same diffusion sensitizing gradients, but in a different order, should yield equivalent information, even though it permutes the temporal order. To the best of our knowledge, no codec has been proposed so far that exploits this very specific structure. There has been extensive work on compressed sensing for diffusion MRI (see \cite{Merlet:2013,Tobisch:2018Frontiers} and references therein), but with a focus on reducing measurement time, rather than efficient storage of the measured data.

Recent work has demonstrated the potential of deep learning for lossless compression of 3D medical images \cite{nagoor2020lossless}. Extending this specifically for diffusion MRI is an interesting future direction. However, our PDE-based approach has the advantage of not requiring any training data. Since medical data is a particularly sensitive type of personal data, obtaining diverse large-scale datasets can be difficult, and the potential of model attacks that could cause data leakage is concerning \cite{Nasr:2019,marwood2018representing}.

\section{Proposed Lossless Codec}
\label{sec:codec}

Traditional PDE-based image compression \cite{galic2008image,10.1007/978-3-030-56215-1_5,jumakulyyev2021lossless,schmaltz2014understanding} performs inpainting in image space, which relies on piecewise smoothness of the image. A key contribution of our current work is to additionally exploit the smoothness in q-space. As it can be seen in Figure~\ref{fig:magneticField-gradient-diffusion}, dMRI signals that are measured with similar gradient directions are correlated.

Our codec uses a spatial PDE for the first few volumes, which is described in more detail in Section~\ref{sec:spatial-codec}. Once sufficiently many samples are available so that a q-space PDE, described in Section~\ref{sec:qSpace-pdePrediction}, produces stronger compression than the spatial PDE, we switch to it.

The q-space PDE assumes that all volumes are in correct spatial alignment, which might be violated in practice due to subject motion. For this reason, our codec includes a mechanism for motion compensation, described in Section~\ref{sec:alignment}. Our overall compressed file format is specified in Section~\ref{sec:coding}.

\subsection{Lossless 3D Spatial Codec}
\label{sec:spatial-codec}
The initial few volumes are compressed with an image space PDE-based codec that follows our recent conference paper \cite{jumakulyyev2021lossless}. To make our current work self-contained, we briefly summarize the most relevant points, focusing on the forward, i.e., encoding direction. The decoding process just mirrors the respective steps. The codec is composed of three main parts: Data sparsification (initial mask selection), prediction (iterative reconstruction), and residual coding.

\textit{Initial Mask Selection:} As an initial mask, our codec simply stores voxel intensities on a sparse regular grid. More specifically, for a given 3D input image of size  $n_1 \times n_2 \times n_3$, the initial mask is chosen as a hexahedral grid consisting of voxels $(4i_1, 4i_2, 4i_3)$, where $i_j \in \{0, 1, \ldots , \lfloor \frac{(n_j-1)}{4} \rfloor \}$, $j \in {1,2,3}$.

Most lossy PDE-based codecs select a mask adaptively \cite{galic2008image,schmaltz2014understanding}, which better preserves important image features such as edges and corners \cite{mainberger2011edge}. However, this introduces the need to store the locations of the selected pixels, which can be avoided by the use of fixed grids \cite{hoffmann2013compression,peter2019fast}. In the context of lossless compression, we achieved higher compression rates by combining the latter strategy with iterative reconstruction.

\textit{Iterative Reconstruction:} Making PDE-based compression lossless requires coding the differences between the original image and the PDE-based reconstruction, and is beneficial in terms of compression rates to the extent that those residuals are more compressible than the original image. In general, residuals become more compressible the more accurate the reconstruction is. Therefore, the overall compression rate can be increased by iteratively coding residuals of some pixels, and refining the remaining ones based on them.

Our previous work \cite{jumakulyyev2021lossless} explored different iterative schemes. The variant that is used here codes the residuals in all remaining face-connected neighbors of the current mask voxels, i.e., up to six voxels per mask voxel. Those neighbors become part of the mask for the next iteration, and the process continues until all voxels have been coded.

Among the PDEs that have been explored for inpainting, we currently consider the two that worked best in \cite{jumakulyyev2021lossless}, i.e., traditional edge-enhancing diffusion (EED) \cite{weickert1998anisotropic} and our recent fourth-order generalization (FOEED) \cite{10.1007/978-3-030-56215-1_5}.

\textit{Residual Coding:} Residuals are computed in modular arithmetic, so that they can be represented as unsigned integers. The final compression of the initial mask and the residuals is either done via a Huffman entropy encoder or the Deflate algorithm, depending on which gives the smaller output file size. 

In cases where medical images contain a substantial amount of empty space, e.g., a background region with exactly zero image intensity, our previous work \cite{jumakulyyev2021lossless} found that coding it separately using run length encoding (RLE) can provide an additional benefit. Unfortunately, in dMRI, the background is perturbed by measurement noise, which renders this approach ineffective. Therefore, our current work does not include any dedicated empty space coding.

\begin{figure}[tbp]
  \centering
  \includegraphics[height=.34\textheight, width=0.99\linewidth]{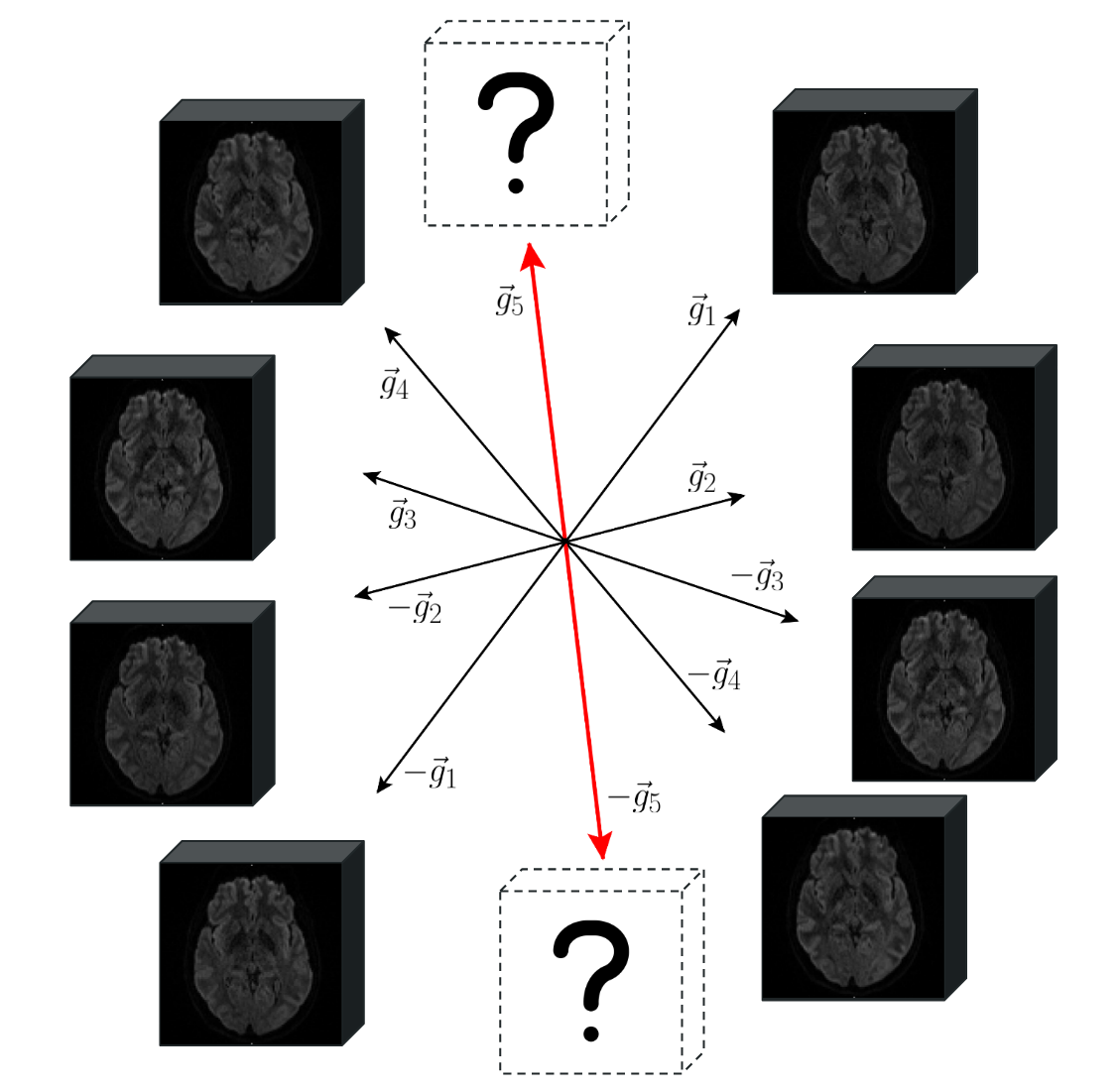}
  \centering
  \caption{\label{fig:q_space_inpaintings}%
    An example of $q$-space inpainting to predict a diffusion-weighted volume in gradient direction $\vec g_5$ (red double arrow) based on four known volumes, illustrated as filled volumes in directions $\vec g_i$ (black arrows).}
\end{figure}

\subsection{PDE-based q-Space Inpainting}
\label{sec:qSpace-pdePrediction}

The general idea of q-space inpainting is illustrated in Figure~\ref{fig:q_space_inpaintings}: Once a certain number of diffusion-weighted images with different gradient directions are known, we can use them to predict images that correspond to a new direction. This happens at the voxel level, so that the prediction at a given location is entirely determined by values at the same location in the known images.

This can be understood as ``flipping'' the setup from Section~\ref{sec:spatial-codec}, where the mask consisted of pixel locations, and the inpainting was repeated with an identical mask for each channel. Instead, the mask now specifies the known channels, and inpainting is repeated for each voxel in the volume.
  
\subsubsection{Compressing Diffusion-Weighted Images}
\label{sec:compressing-dwis}

Since we assume that diffusion-weighted measurements are on spherical shells in q-space (Section~\ref{subsec:dMRI}), we inpaint with second-order linear homogeneous (LH) diffusion
\begin{equation}
  \label{eq:laplace}
  \partial_t u=\Delta u
\end{equation}
or fourth-order biharmonic (BH) smoothing
\begin{equation}
  \label{eq:biharmonic}
  \partial_t u=-\Delta^2 u
\end{equation}
on the sphere, where $\Delta$ is the Laplace-Beltrami operator.

\begin{figure}[tbp]
  \centering
  \includegraphics[height=.19\textheight, width=.492\linewidth]{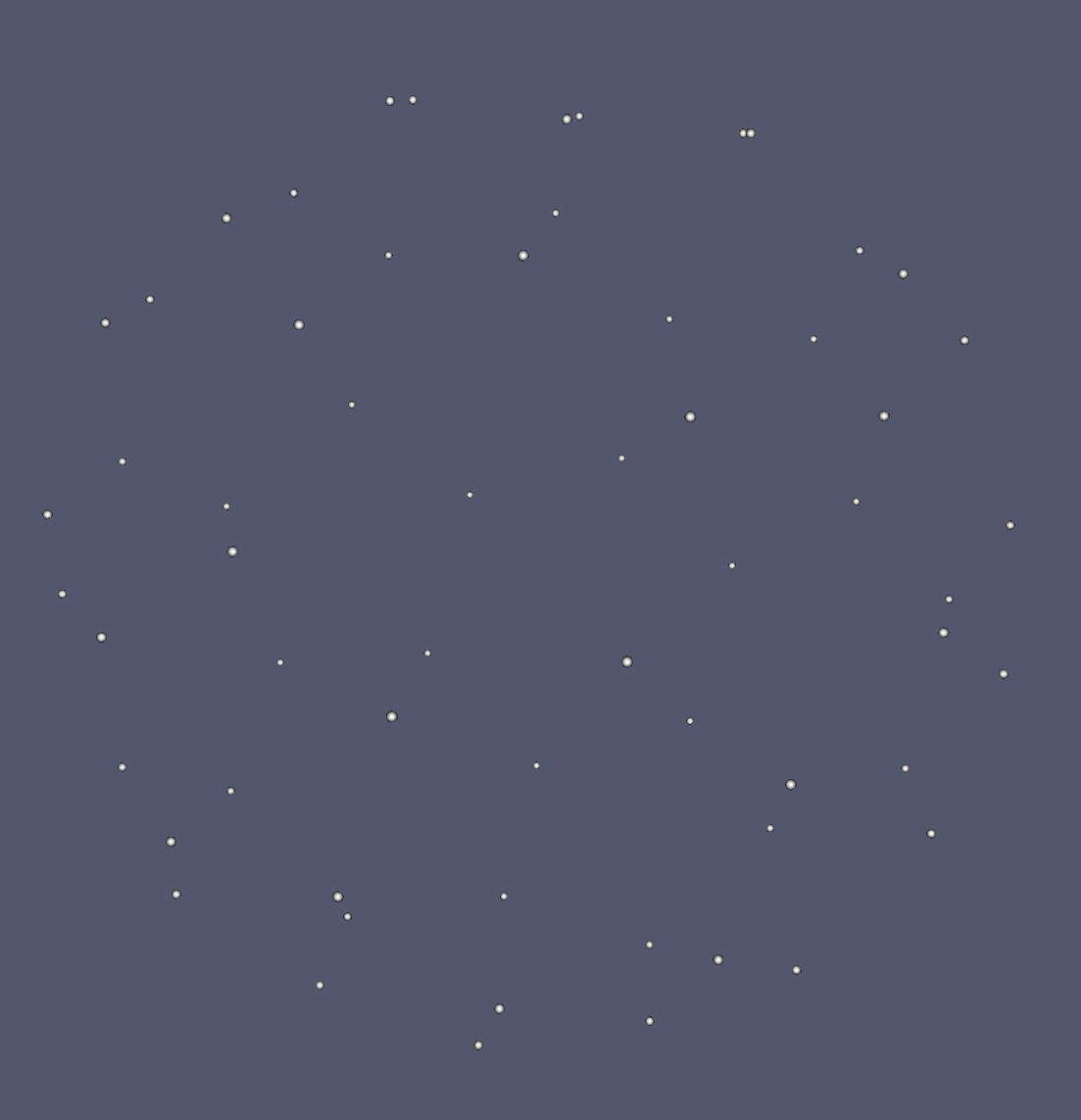}
  \includegraphics[height=.19\textheight, width=.492\linewidth]{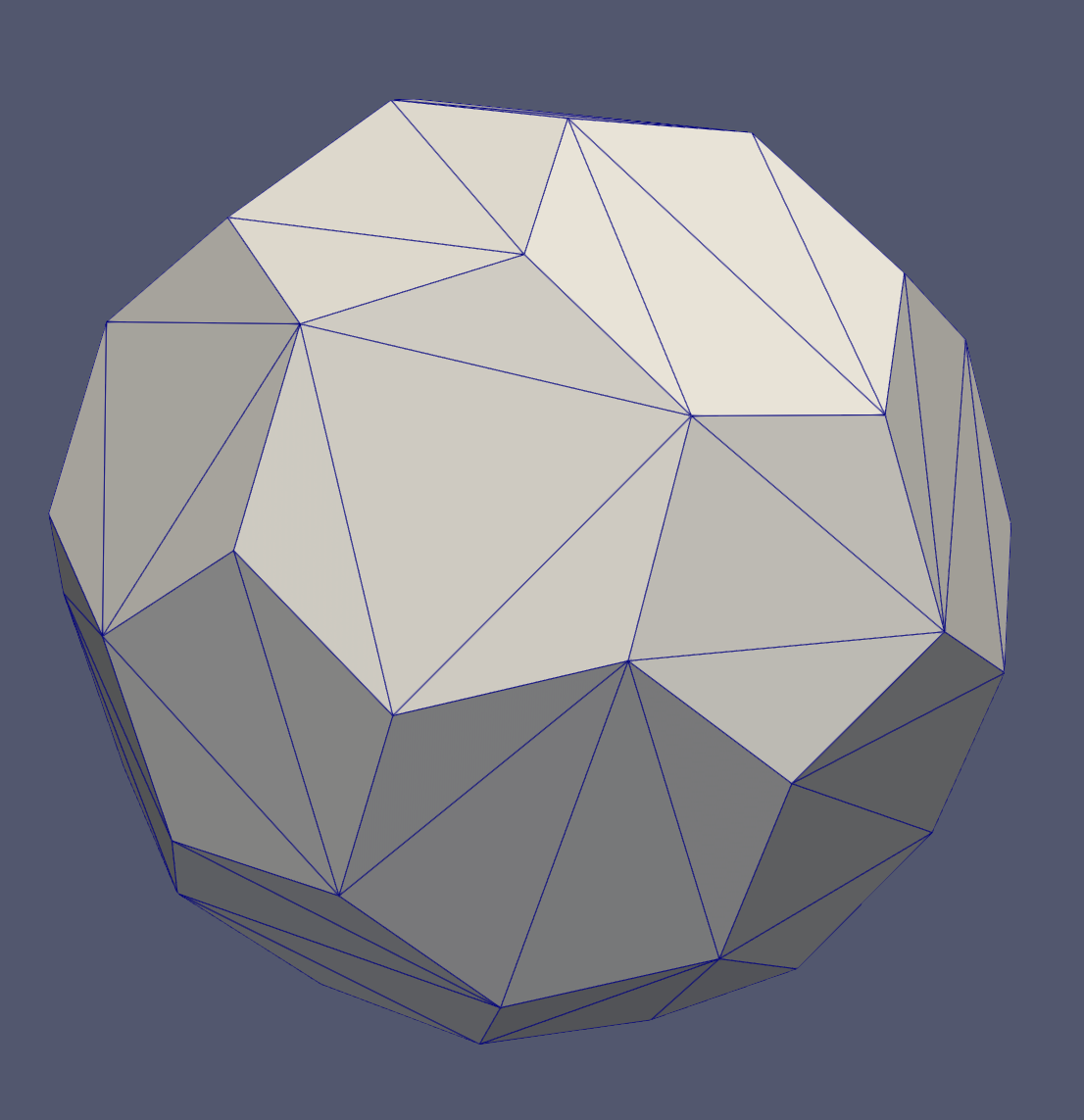}
  \hfill \mbox{}      
  \centering
  \caption{\label{fig:qSpaceSampling}%
    The $q$ space sampling of the dMRI data used in our experiments (left), and the resulting triangulation that is used within the Finite Element Method.}
\end{figure}

Given that our samples do not form a regular grid, we numerically solve these equations using Finite Element Methods (FEM) \cite{Chizhov:2021,logg2012automated}. For this, we first construct a 3D Delaunay tessellation from the set of all gradient vectors $\mathbf{g}_i$ and their antipodal counterparts $-\mathbf{g}_i$, and then extract a triangular surface mesh from it. Figure~\ref{fig:qSpaceSampling} shows an example of the given vectors (left), and the resulting triangular mesh (right). 

Similar to PDE-based inpainting in the image domain, we fix the known values by imposing Dirichlet boundary conditions at the vertices corresponding to the previously coded diffusion-weighted images, again accounting for antipodal symmetry so that each known image determines the values of two vertices. Once a steady state has been reached, the values at locations corresponding to diffusion-weighted images that are yet to be coded can serve as predictions. Similar as before, we compute residuals with respect to those predictions in modular arithmetic, and apply Huffman coding or Deflate to them.

We found that, once a sufficient number of diffusion weighted images are available as a basis of q-space inpainting, its residuals become more compressible than those from iterative image space inpainting. Our codec adaptively determines a suitable point for switching from spatial to q-space predictions. After the first diffusion-weighted volume, it starts comparing the sizes of compressing subsequent volumes with the spatial codec (Section~\ref{sec:spatial-codec}) to the size when using q-space inpainting and switches to it on the first volume where it is beneficial. To limit computational effort, the spatial codec is no longer tried for subsequent volumes.

\subsubsection{Accelerated Computation}
\label{sec:faster-computation}

Since q-space inpainting happens at a voxel level, it should be repeated for each voxel of the 3D image. However, the computational cost of running the FEM solver for each voxel separately is extremely high. Fortunately, linearity of the PDEs and the fact that the Dirichlet boundary conditions are imposed on the same vertices for each voxel permit a significant speedup.

Formally, we can consider one time step of numerically solving Equation~(\ref{eq:laplace}) or~(\ref{eq:biharmonic}), at time $t$, as applying a discrete linear differential operator $\mathbf{D}$, which is determined by the vertices and connectivity of our triangular mesh, on a discrete input $\mathbf{u}^{(t)}\in\mathbb{R}^c$,
\begin{equation}
	\label{eq:LinearDiffOper}
	\mathbf{u}^{(t+1)}=\mathbf{D} \left[\mathbf{u}^{(t)}\right],
\end{equation}
where $c$ is the number of channels (q-space samples per voxel). The boundary conditions ensure that $u_{k_j}^{(t+1)}=u_{k_j}^{(t)}$ at positions $k_j$ that correspond to the fixed (previously coded) channels.

The inpainting result is obtained as the fixed point $\mathbf{u}^{(\mathrm{FP})}$ as $t\rightarrow\infty$. It can be approximated by iterating $\mathbf{D}$ a sufficient number of times, resulting in an operator $\mathbf{D}_\mathrm{FP}$ that directly maps
\begin{equation}
  \label{eq:FixedPoint}
  \mathbf{u}^{(\mathrm{FP})}=\mathbf{D}_\mathrm{FP} \left[\mathbf{u}^{(0)}\right].
\end{equation}

$\mathbf{D}_\mathrm{FP}$ is still linear, and we observe that its
kernel is the subspace corresponding to the unknown q-space samples,
so that their initialization in $\mathbf{u}^{(0)}$ does not influence the
steady state \cite{peter2015beyond}. Therefore, we can rewrite
Equation~\ref{eq:FixedPoint} as
\begin{equation}
  \label{eq:LinearDiffOper_seperated}
  \begin{aligned}
      \mathbf{D}_\mathrm{FP} \left[\mathbf{u}^{(0)}\right]=
  &u^{(0)}_{k_1} \mathbf{D}_\mathrm{FP} [\mathbf{e}_{k_1}] + u^{(0)}_{k_2} \mathbf{D}_\mathrm{FP} [\mathbf{e}_{k_2}]\\& + \ldots + u^{(0)}_{k_n} \mathbf{D}_\mathrm{FP} [\mathbf{e}_{k_n}],
  \end{aligned}
\end{equation}
where $\mathbf{e}_{k_j}$ are the indicator vectors of the known samples $k_j$.

In other words, by computing $\mathbf{w}_{k_j}=\mathbf{D}_\mathrm{FP} [\mathbf{e}_{k_j}]$, we can obtain weight vectors that specify how the known values are combined to predict the unknown ones. They are analogous to the ``inpainting echoes'' that have been computed in previous work \cite{mainberger2011optimising} for the purpose of optimizing tonal data. Omitting the irrelevant initialization of the unknown values from the input $\mathbf{u}^{(0)}$, and the known values from the output $\mathbf{u}^{(\mathrm{FP})}$ yields a weight matrix $\mathbf{W}$ of shape $m\times n$ for $n$ known and $m$ unknown values.

We compute the coefficients of $\mathbf{W}$ by running the FEM $n$ times. In the $j$th run, we set the value corresponding to $k_j$ to one, all remaining values to zero. After numerically solving the Laplace or Biharmonic PDE, the values at the $m$ unique vertices that correspond to unknown DWIs yield the $j$th column of $\mathbf{W}$.

Finally, $\mathbf{W}$ allows us to make efficient predictions in each voxel, by simply multiplying it to a vector that contains the intensities in that voxel from the previously coded diffusion gradients.

\subsubsection{Implementation Details and Running Times}
\label{sec:implementation-details}

We numerically solve Equations~(\ref{eq:laplace}) and~(\ref{eq:biharmonic}) via the open-source FEM solver package FEniCSx \cite{logg2012automated}. For implementation details, we refer to its tutorials \cite{langtangen2017solving}. Applying this solver to each voxel of a $104 \times 104 \times 72$ volume takes close to two and four hours, respectively, for LH and BH PDEs on a single 3.3\,GHz CPU core.

The acceleration from the previous section reduces this to only 1.6\,s and 2.4\,s per volume, respectively. This includes the time for building a Delaunay tessellation, which is computed with SciPy \cite{2020SciPy-NMeth}, and extracting a surface mesh from it using the \texttt{BoundaryMesh} method from FEniCSx.

\subsubsection{Compressing $b=0$ Images}
\label{sec:compressing-b0}

\begin{figure}
  \centering
  \includegraphics[height=.29\textheight, width=0.99\linewidth]{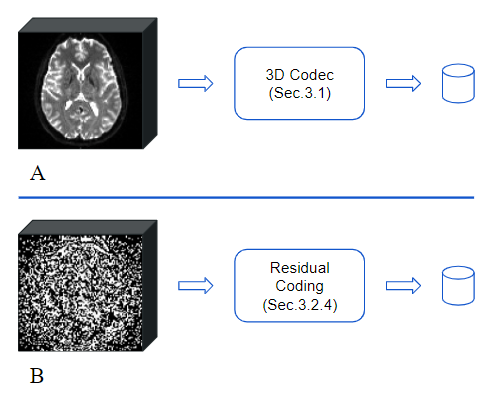}
  \centering
  \caption{\label{fig:b0_init_b700_compression}%
   Top (A): Compression of the first $b=0$ volume using the recently proposed lossless 3D codec \cite{jumakulyyev2021lossless}.
   Bottom (B): Compression of the remaining $b=0$ volumes using residuals in modular arithmethic. Residuals are taken with respect to the first $b=0$ volume, after motion correction.}	
\end{figure}

Our general approach simplifies for unweighted volumes with $b=0$. Again, the first of them is compressed using the spatial codec. If multiple $b=0$ images were acquired to increase the signal-to-noise ratio, our codec compresses the remaining ones by taking the respective residuals with respect to the first $b=0$ volume, as illustrated in Figure~\ref{fig:b0_init_b700_compression}.

\subsection{Motion Correction}
\label{sec:alignment}
Subject motion commonly occurs during the lengthy dMRI acquisitions, and is typically accounted for by applying motion correction based on image registration \cite{jenkinson2002improved}. We also include this step in our codec, since inpainting in q-space requires a correct spatial alignment of all 3D volumes so that predictions are based on information from the same location within the brain.

We implement motion correction as follows:
\begin{enumerate}[1.]
\item We perform affine registration of each volume to the same $b=0$ volume, which is used as a common reference. This yields a transformation matrix $\mathbf{T}_{X \rightarrow b_0}$ which aligns DWI volume $X$ to the $b=0$ reference.
\item When predicting a DWI volume $P$, we transform all known volumes $X$ via the affine transformation $\mathbf{T}_{P \rightarrow b_0}^{-1} \circ \mathbf{T}_{X \rightarrow b_0}$, which can be computed from the transformations in Step~1.
\item In addition to resampling each known volume $X$, we re-orient its diffusion gradient direction $\mathbf{g}_X$ according to the rotational part of the transformation in Step~2. Omitting this step would lead to incorrect relative orientations of diffusion gradient directions \cite{leemans2009b}, which could again reduce accuracy of q-space inpainting.
\end{enumerate}

Transforming images via a common reference allows us to align them without having to perform image registration on all pairs of volumes. This saves considerable computational effort. Combining the two transformations and applying them in a single step also reduces computational effort, and simultaneously reduces image blurring compared to a two-stage implementation that would involve interpolating twice.

In addition to the computational expense, motion correction incurs the
cost of having to store the affine matrices $\mathbf{T}_{X \rightarrow b_0}$
along with the compressed data. Experimental results in
Section~\ref{subsec:results-motion-correction} will demonstrate that
this storage cost is outweighed by the increase in compression rate when q-space inpainting properly accounts for motion.

\begin{figure}
  \centering
  \includegraphics[height=.14\textheight, width=.99\linewidth]{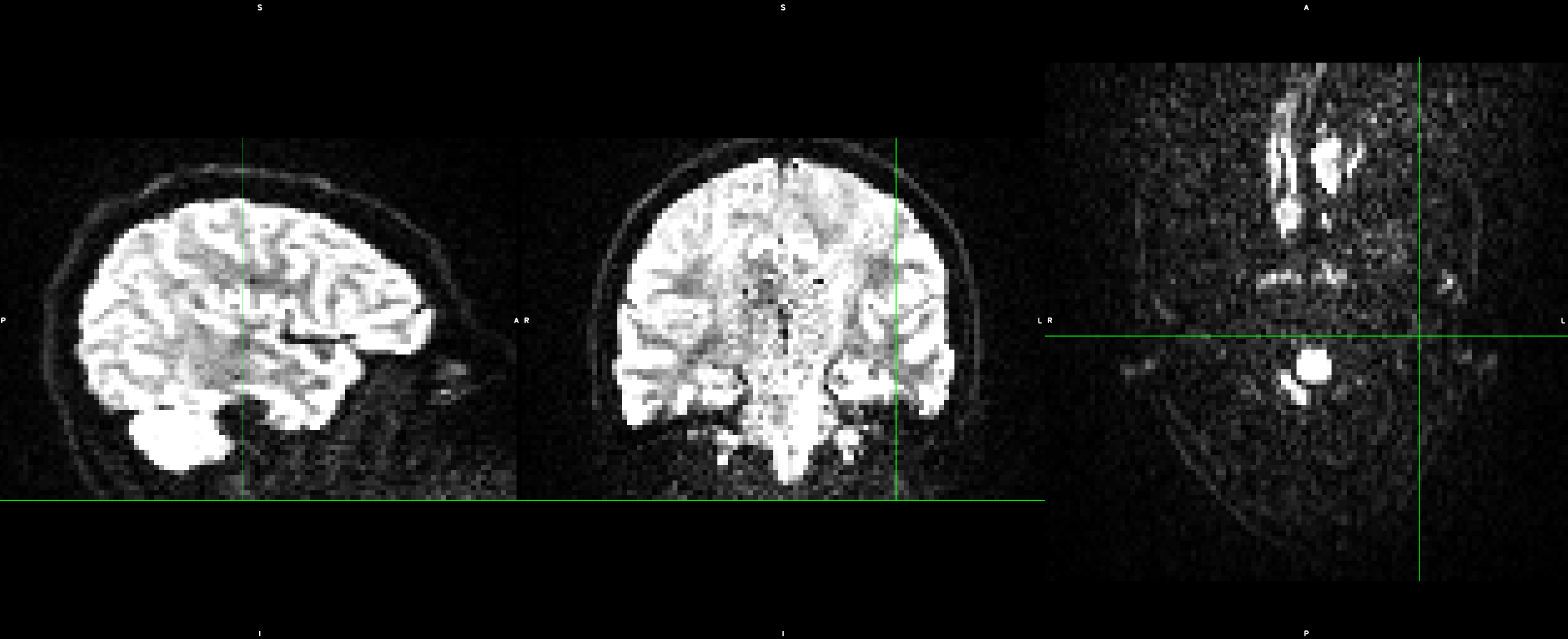}
  \hfill
  \includegraphics[height=.14\textheight, width=.99\linewidth]{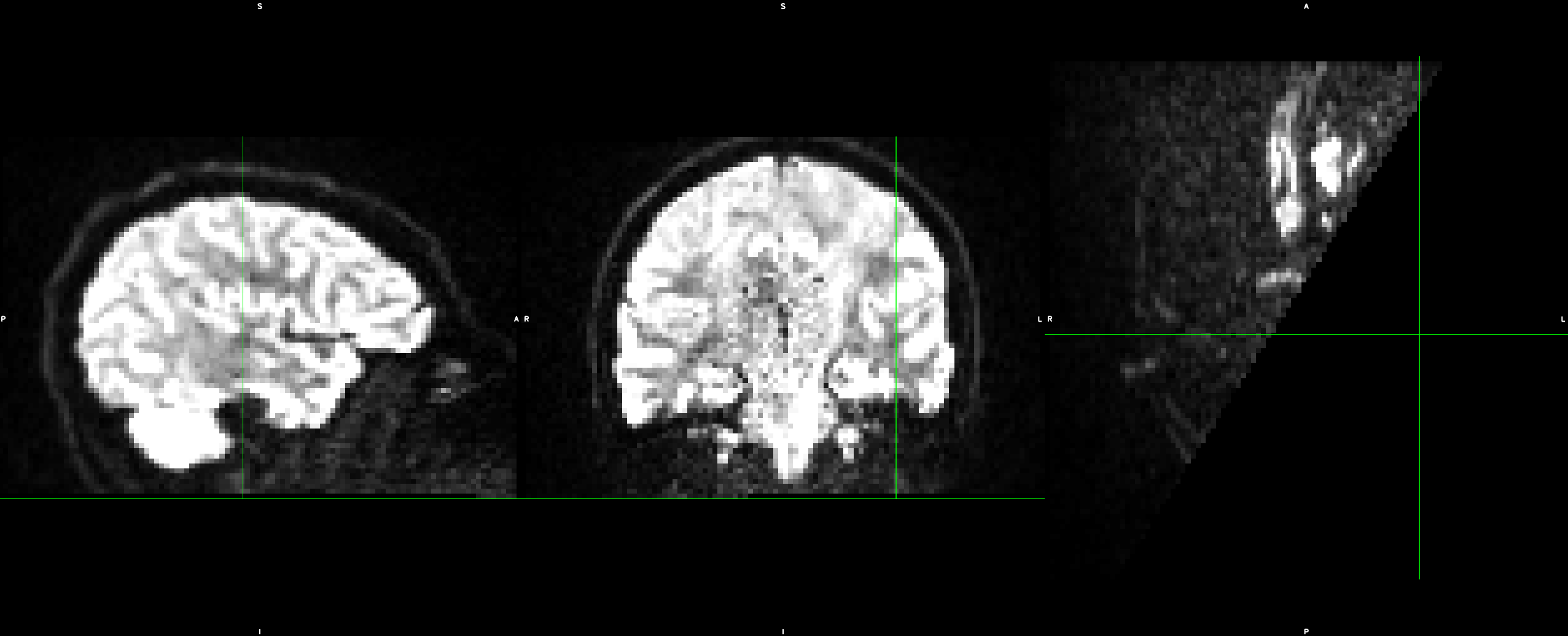}
  \hfill
  \includegraphics[height=.14\textheight, width=.99\linewidth]{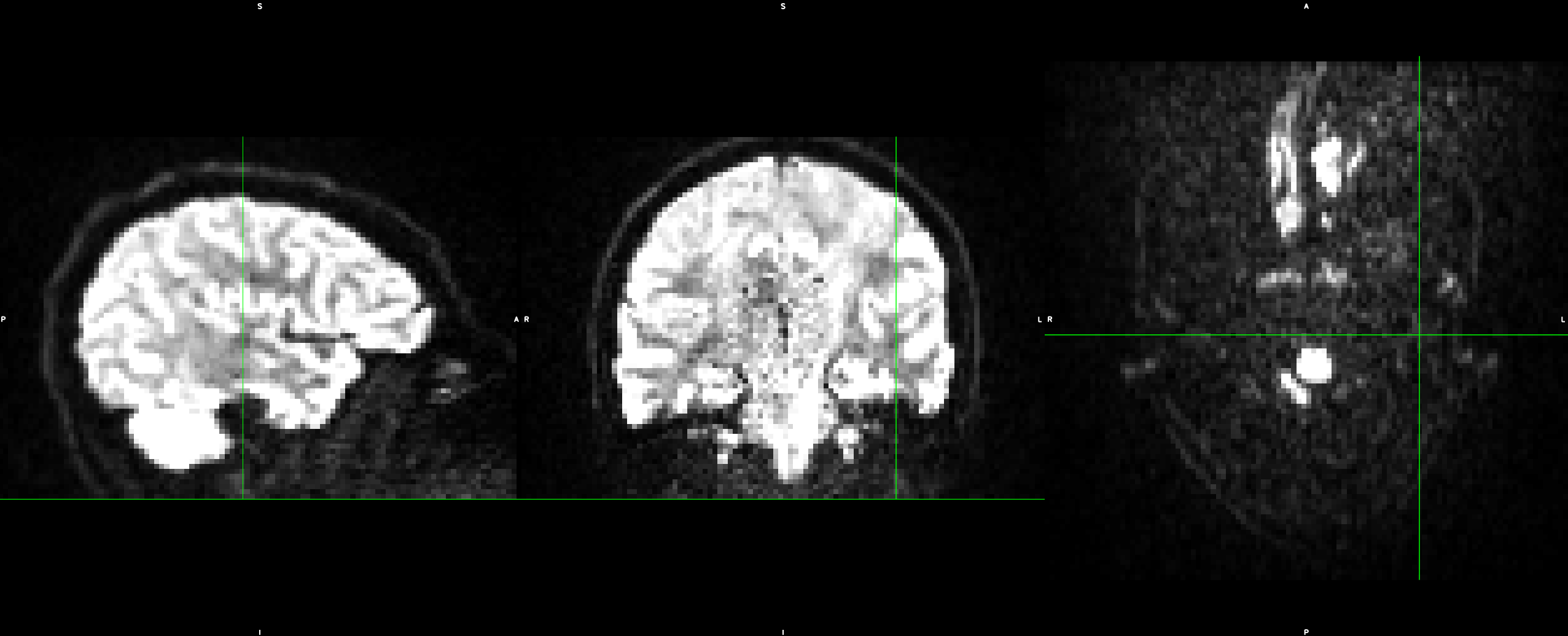}
  \centering
  \caption{\label{fig:boundary_effects}%
    Boundary effects in volume alignment. Top: Original DWI volume. Center and Bottom: Motion corrected with zero padding and nearest neighbor extrapolation, respectively.}
\end{figure}

Subject movement correction and B-matrix reorientation are done using the freely available FSL tools \cite{jenkinson2002improved} and the DIPY imaging library \cite{garyfallidis2014dipy}, respectively. A practically relevant implementation detail concerns boundary effects. As illustrated in Figure~\ref{fig:boundary_effects}, missing information can enter the field of view when applying image transformations. We found that q-space inpainting near the boundary of the domain works more reliably if we resolve these cases with nearest neighbor extrapolation, rather than with zero padding.

\begin{figure*}
  \centering
  \hfill
  \includegraphics[width=0.475\linewidth]{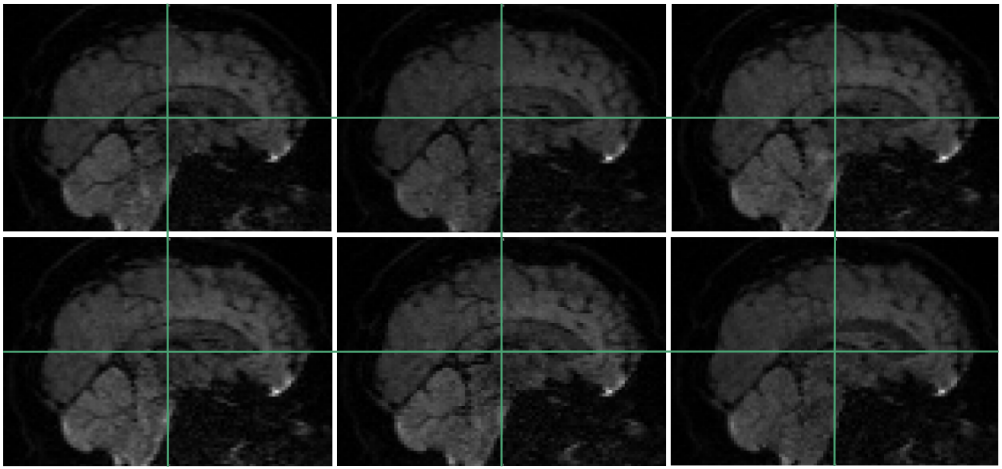}
  \hfill \hfill
  \includegraphics[width=0.475\linewidth]{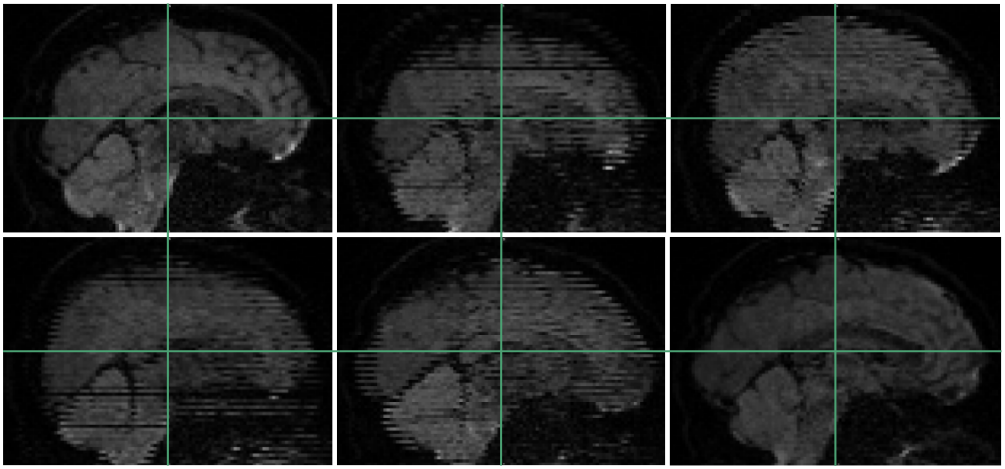}
  \hfill
  \caption{\label{fig:dMRI_data}%
    Example images from our two dMRI datasets, without deliberate head motion (left) and with strong motion artifacts (right). In each case, six corresponding sagittal slices from different diffusion weighted images (DWIs) are shown. Note that subject motion leads to spatial misalignments between DWIs, but also to artifacts within individual images.}
\end{figure*}

\subsection{Compressed File Format}
\label{sec:coding}
In our current implementation, the relevant data is spread over multiple files whose sizes are added when computing compression rates.

The volumes that are compressed with the 3D lossless codec (Section~\ref{sec:spatial-codec}) are stored with the same header as in \cite{jumakulyyev2021lossless}. Stated briefly, it contains the original minimum and maximum voxel values (4~bytes), sizes of the compressed data streams for zero voxel binary mask and mask intensities (8~bytes), the diffusivity contrast parameter (4~bytes), the type of PDE (2~bits), the dilation mode (1~bit), and the types of encoding for mask intensities and residuals (2~bits).

For each volume that is compressed with q-space inpainting, the header contains the original minimum and maximum voxel values (4~bytes), the type of PDE (2~bits), the type of encoding for the residuals (1~bit), and the volume number in the original order (2~bytes).

Mask and residual values themselves are stored after compression with pure Huffman coding or Deflate, depending on what gave a smaller file size.

In addition, we store the NIfTI header (348 bytes) as well as files containing b values and gradient vectors in their original ASCII formats. For simplicity, affine transformations for motion correction are also kept in the ASCII format generated by FSL FLIRT \cite{jenkinson2002improved}.

\section{Results and Discussion}
\label{sec:results}

\subsection{Data}
\label{subsec:data}

We evaluate our codec on two dMRI datasets that were made publicly available by Koch et al.\ \cite{koch2019shore}, and are specifically suited to investigate the impact of subject motion compensation. Both datasets have been collected from the same subject (male, 34 years) in the same scanner, a 3T MAGNETOM Skyra (Siemens Healthcare, Erlangen, Germany), with an identical measurement sequence.
For the first scan, the subject received the usual instruction of staying as still as possible during the acquisition. For the second scan, the subject was asked to move his head, to deliberately introduce motion artifacts.

From these datasets, we use the five non-diffusion weighted ($b=0$) MRI scans each, as well as 30~diffusion weighted images ($b=700\,\mathrm{s}/\mathrm{mm}^2$, diffusion gradient duration $\delta=334\,\mathrm{ms}$, spacing $\tau=445\,\mathrm{ms}$). Each image consists of $104 \times 104 \times 72$ voxels with a resolution of $2\times 2\times 2\,\mathrm{mm}^3$. The data, and the effects of subject motion, are illustrated in Figure~\ref{fig:dMRI_data}.

\subsection{DTI Baseline}
\label{subsec:DTI}

We compare the signal predictions from our q-space PDE to a simple baseline, which is derived from the Diffusion Tensor Imaging (DTI) model. DTI is widely used in practice, due to its relative simplicity and modesty in terms of scanner hardware and measurement time.

It rests on the assumption that the diffusion propagator $\bar{P}(\mathbf{R},\tau)$ is a zero-mean Gaussian whose covariance matrix is proportional to the diffusion tensor $\mathbf{D}$, a symmetric positive definite $3\times 3$ matrix that characterizes the local diffusion \cite{Basser:1994}. The signal model in DTI relates the diffusion-weighted signal $S(\hat{\mathbf{g}},b)$ for a given b-value and gradient vector direction $\hat{\mathbf{g}}=\mathbf{g}/\|\mathbf{g}\|$ to the unweighted signal $S_0$ according to
\begin{equation}
  \label{eq:dti-signal-equation}
 S(\hat{\mathbf{g}},b)=S_0\mathrm{e}^{-b\hat{\mathbf{g}}^T\mathbf{D}\hat{\mathbf{g}}}.
\end{equation}

\begin{table*}
  \caption{Compressed file sizes from separate PDE-based compression of each 3D scan (baseline), from different variants of our proposed lossless codec, as well as from GZIP and lossless codecs from the JPEG family. For hybrid codecs, the split indicates the number of volumes coded with q-space or spatial inpainting, respectively.}\label{qPDEs_3DCodec}
\begin{center}
\begin{tabular}{lccccccc}
  \toprule%
  & \multicolumn{3}{c}{Scan~1: No deliberate motion} & & \multicolumn{3}{c}{Scan~2: Strong head motion}\\
  \cmidrule{2-4}
  \cmidrule{6-8}
  Codec Variant & Split &\vtop{\hbox{\strut Size}\hbox{\strut (bytes)}} & \vtop{\hbox{\strut Over}\hbox{\strut R-IEED-1}} & &Split &\vtop{\hbox{\strut Size}\hbox{\strut (bytes)}} & \vtop{\hbox{\strut Over}\hbox{\strut R-IEED-1}}\\
  \midrule
  R-IEED-1 & & 16022666 & - & & &16082537 & - \\
  R-IFOEED-1 & & 15955826 & +0.42\% & & &16019913 & +0.39\% \\
\midrule  
qLH $\circ$ R-IFOEED-1 & 27/4 & 14984472 & +6.50\% & & 26/5 & 15570493 & +3.18\% \\
qLH $\circ$ R-IEED-1 & 27/4 & 14991732 & +6.43\% & & 26/5 & 15578604 & +3.13\% \\
qBH $\circ$ R-IFOEED-1 & 27/4 & 15032384 & +6.18\% & &26/5 & 15681354 & +2.50\% \\
qBH $\circ$ R-IEED-1 & 27/4 & 15039644 & +6.14\% & &26/5 & 15689465 & +2.44\% \\
DTI $\circ$ R-IFOEED-1 & 24/7 & 15099213 & +5.76\% & &24/7 & 15854216 & +1.42\%\\
  DTI $\circ$ R-IEED-1 & 24/7 & 15108244 & +5.71\% & &24/7 & 15861176 & +1.38\%\\
  \midrule
GZIP & & 21841701 & -36.32\% & & &21819641 & -35.67\% \\
JPEG &  & 17885953 & -11.63\% & & &17905933 & -11.34\% \\
JPEG-LS & & 17921807 & -11.85\% & & &17893931 & -11.26\% \\
JPEG2000 & & 15993453 & +0.18\% & & &15980005 & +0.64\% \\
\botrule
\end{tabular}
\end{center}
\end{table*}

Fitting $\mathbf{D}$ requires at least one reference MR image $S_0$, plus diffusion-weighted images in at least six different directions, which are usually taken with a fixed non-zero b-value. Equation~(\ref{eq:dti-signal-equation}) can then be used to predict the diffusion-weighted signal in any desired direction. In our experiments, we compare our PDE-based to DTI-based predictions that account for the same set of known measurements.

\subsection{Comparing Lossless Codecs for Diffusion MRI}
\label{subsec:lossless-codecs}

A comparison of file sizes that can be achieved on our two test datasets with different lossless codecs is provided in Table~\ref{qPDEs_3DCodec}. As a baseline, the first two rows show results from coding each 3D volume independently with our recently proposed PDE-based codec \cite{jumakulyyev2021lossless}, using second-order (R-IEED-1) and fourth-order anisotropic diffusion (R-IFOEED-1). Additional savings of other codecs with respect to R-IEED-1 are given in percent.

The second block in Table~\ref{qPDEs_3DCodec} shows results from several variants of our proposed new codec, which adaptively combines inpainting in q-space and image space. Highest compression rates were achieved when combining linear homogeneous (LH) diffusion in q-space with R-IFOEED-1 in image space, closely followed by R-IEED-1. Biharmonic (BH) diffusion in q-space also produced useful, but slightly weaker results.

Both q-space diffusion approaches achieved better compression than predictions from DTI (Section~\ref{subsec:DTI}). This could be due to the fact that the quadratic model of diffusivities in Equation~(\ref{eq:dti-signal-equation}) is known to be an oversimplification in many parts of the brain \cite{Alexander:2002}, and the PDE-based approaches provide more flexibility.

DTI requires independent coding of at least seven 3D volumes, which led us to fix this split in our experiments. PDE-based imputation makes it possible to switch to q-space inpainting earlier, and our adaptive selection does so after four volumes in the low-motion data, after five volumes in the data with strong motion.

Switching to q-space inpainting also speeds up our codec. Our implementation of R-IEED-1 and R-IFOEED-1 requires approximately 478\,s and 6185\,s, respectively, for one volume on a single 3.3\,GHz CPU core. Even though it would be possible to further optimize this, exploiting linearity in qLH and qBH, as described in Section~\ref{sec:qSpace-pdePrediction}, significantly lowers the intrinsic computational complexity, so that even a straightforward implementation only requires 1.64\,s and 2.4\,s per volume, respectively.

\begin{table*}
\begin{center}
  \caption{Compressed file sizes when omitting motion compensation, and the relative benefit from motion correction.}\label{qPDEs_3DCodec-data_without_motion_artifacts-motion_compensation_benefit}
\begin{tabular}{lccccccc}
  \toprule%
  & \multicolumn{3}{c}{Scan~1: No deliberate motion} & & \multicolumn{3}{c}{Scan~2: Strong head motion}\\
  \cmidrule{2-4}
  \cmidrule{6-8}
Codec Variant  & Split & \vtop{\hbox{\strut Without}\hbox{\strut Correction}} & Benefit & & Split & \vtop{\hbox{\strut Without}\hbox{\strut Correction}} & Benefit\\
\midrule
qLH $\circ$ R-IFOEED-1 & 27/4 & 15113407 & +0.85\% &  & 16/15 & 16456557 & +5.38\%\\
qLH $\circ$ R-IEED-1 & 27/4 & 15122438 & +0.86\% &  & 16/15 & 16478088 & +5.46\%\\
qBH $\circ$ R-IFOEED-1 & 27/4 & 15302805 & +1.77\% &  & 16/15 & 16582150 & +5.43\%\\
qBH $\circ$ R-IEED-1 & 27/4 & 15311836 & +1.78\% &  & 16/15 & 16603681 & +5.51\%\\
DTI $\circ$ R-IFOEED-1 & 24/7 & 15396194 & +1.93\% &  & 24/7 & 16946648 & +6.45\%\\
DTI $\circ$ R-IEED-1 & 24/7 & 15405225 & +1.93\% &  & 24/7 & 16955494 & +6.45\%\\
\botrule
\end{tabular}
\footnotetext{Compressed file size comparision of our proposed qLH- and qBH-based codecs with and without motion correction. Second column: Compression file sizes (in bytes) without motion compensation. Third column: Improvement percentage when considering motion compensation. The first six codec variants are from dMRI with no noticible motion artifact compression and the last six ones are from compression dMRI dataset with heavy subject motion.}
\end{center}
\end{table*}

It can be seen in Figure~\ref{fig:dMRI_data} that subject motion during different phases of the acquisition leads to different types of artifacts. Results in Table~\ref{qPDEs_3DCodec} include the motion correction described in Section~\ref{sec:alignment}, which compensates spatial misalignments of different scans. However, motion can also lead to signal dropouts or to distortions within scans, which our current codec does not explicitly account for. This explains why q-space inpainting is less effective on the second as compared to the first scan. However, even on this challenging dataset that exhibits unusually strong artifacts, q-space inpainting still provides a benefit compared to all other alternatives.

Finally, Table~\ref{qPDEs_3DCodec} shows results from several other lossless codecs for comparison. GZIP is most widely used in practice, but the resulting files are more than 35\% larger than those from our proposed codec. Among the lossless codecs from the JPEG family, JPEG2000 is the only one that outperforms R-IEED-1 for per-volume compression, and only by a small margin. Our new hybrid methods that combine image space and q-space inpainting always performed best.

\subsection{Benefit from Motion Correction}
\label{subsec:results-motion-correction}

Table~\ref{qPDEs_3DCodec-data_without_motion_artifacts-motion_compensation_benefit} investigates the benefit of motion correction (Section~\ref{sec:alignment}) by showing file sizes when removing motion correction from our codec, and comparing the results to ones with motion correction (Table~\ref{qPDEs_3DCodec}), indicating the benefit in percent.

Even on the first scan, in which the subject tried to keep his head still, compensating for small involuntary movements yields a slight benefit. The effect is largest when imputing via qBH and DTI. This might be explained by the fact that qLH satisfies the min-max principle, which makes it more robust against inaccuracies in its inputs, and provides another argument in its favor.

When strong head motion is present (second scan), restoring a correct voxel alignment via motion correction becomes essential for q-space inpainting. Without it, the switch to q-space imputation happens much later, and the overall file size is larger than when coding each volume independently. This is explained by the fact that our codec always applies difference coding to the $b=0$ images, and that this becomes detrimental when those images are strongly misaligned.

\subsection{Effect of Re-ordering DWIs}
\label{subsec:results-order}

Since q-space imputation relies on the previously (de)coded diffusion weighted images, its accuracy depends on the order in which we process the gradient directions.

\begin{figure}
  \centering
  \includegraphics[height=.167\textheight, width=.492\linewidth]{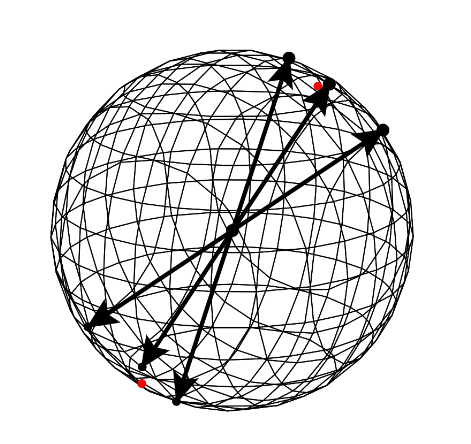}  \includegraphics[height=.167\textheight, width=.492\linewidth]{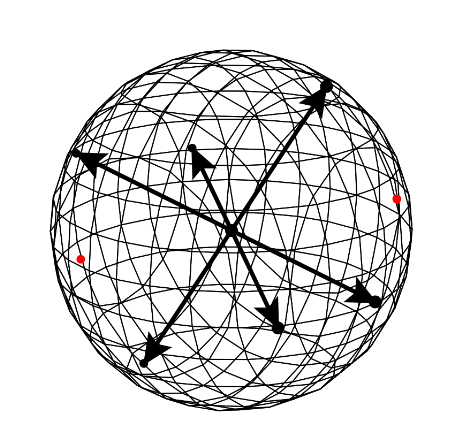}

  \hfill \mbox{}      
  \centering
  \caption{\label{fig:furthest-closest}%
    Given a set of previously coded DWIs, the \emph{closest} strategy (left) selects the volume whose gradient vector has the smallest angular distance from the known ones, to maximize expected prediction accuracy. The \emph{furthest} strategy (right) maximizes the angular distance, aiming for a more uniform coverage of the sphere for subsequent steps. The sketch shows the directions selected in the first three steps as black double arrows, the fourth direction as a red dot.
  }
\end{figure}

Two contradictory greedy strategies are illustrated in Figure~\ref{fig:furthest-closest}: Always selecting the \emph{closest} gradient direction, i.e., the one with the smallest angular distance from the already known ones, can be expected to result in the most accurate prediction, in the same spirit as our spatial codec (Section~\ref{sec:spatial-codec}) iteratively grows a mask of known pixels around an initial set of seed points.

\begin{table}
\begin{center}
  \caption{Compressed file size for scan~1 (without strong motion) when ordering the diffusion-weighted images differently. This affects the accuracy of q-space imputation.}\label{qPDEs_3DCodec-data_without_motion_artifacts-selection_strategy}
\begin{tabular}{lcc}
\toprule%
Codec Variant  & \vtop{\hbox{\strut Closest}\hbox{\strut Selection}} & \vtop{\hbox{\strut Furthest}\hbox{\strut Selection}} \\
\midrule
qLH $\circ$ R-IEED-1 & 15031703 & 14991732\\
qBH $\circ$ R-IEED-1 & 15153333 & 15039644\\
qDTI $\circ$ R-IEED-1 & 15290105 & 15108244\\
\botrule
\end{tabular}
\end{center}
\end{table}

On the other hand, the spatial codec starts with a seed mask that covers the full domain sparsely, but uniformly. Achieving something similar motivates selecting the gradient direction that is \emph{furthest} from any of the known ones. Even though this strategy can be expected to lead to lower accuracy, and therefore to less compressible residuals in the first few iterations, later iterations might benefit from the more uniform coverage of the overall (spherical) domain.

Table~\ref{qPDEs_3DCodec-data_without_motion_artifacts-selection_strategy} presents the effect of these two selection strategies on final file sizes. The results are from the first scan, without strong motion. Overall, greedily selecting the furthest gradient vector gives slightly smaller overall file sizes. Therefore, this is the strategy that we followed in all other experiments.

\section{Conclusion}
\label{sec:conclusion}
In this work, we introduced a PDE-based lossless image compression codec that explicitly exploits both the spatial and the q-space structure in diffusion MRI. To our knowledge, it is the first codec that has been tailored to this type of data. We demonstrated a clear improvement over PDE-based codecs that treat each volume separately, and over other established baselines including GZIP and spatial codecs from the JPEG family.

We evaluated several variants of our codec, and found that q-space predictions with linear homogeneous diffusion permitted the highest compression rates among them. With our proposed method for accelerated computation, it could also be applied at a very reasonable computational cost. We further demonstrated the importance of including motion correction, and propose an efficient implementation that is based on affine image transformations via a common reference. Finally, we found that the order of coding the diffusion-weighted volumes had a relatively minor effect, but that a greedy strategy that strives to cover the sphere as uniformly as possible provides a small benefit.

In the future, one might attempt to replace the switching between image space and q-space inpainting with a PDE that jointly operates on the product space. However, this is likely to substantially increase the computational effort, and introduces the issue of properly balancing image space and q-space diffusion. Similarly, employing nonlinear PDEs for q-space predictions might further increase compression rates, but is likely to cause a high computational cost.

\backmatter
\bmhead{Acknowledgments}
This work has been supported by the German Academic Exchange Service (DAAD).

\bmhead{Conflict of interest}
The authors declare that they have no conflict of interest.



\begin{thebibliography}{72}
\ifx \bisbn   \undefined \def \bisbn  #1{ISBN #1}\fi
\ifx \binits  \undefined \def \binits#1{#1}\fi
\ifx \bauthor  \undefined \def \bauthor#1{#1}\fi
\ifx \batitle  \undefined \def \batitle#1{#1}\fi
\ifx \bjtitle  \undefined \def \bjtitle#1{#1}\fi
\ifx \bvolume  \undefined \def \bvolume#1{\textbf{#1}}\fi
\ifx \byear  \undefined \def \byear#1{#1}\fi
\ifx \bissue  \undefined \def \bissue#1{#1}\fi
\ifx \bfpage  \undefined \def \bfpage#1{#1}\fi
\ifx \blpage  \undefined \def \blpage #1{#1}\fi
\ifx \burl  \undefined \def \burl#1{\textsf{#1}}\fi
\ifx \doiurl  \undefined \def \doiurl#1{\url{https://doi.org/#1}}\fi
\ifx \betal  \undefined \def \betal{\textit{et al.}}\fi
\ifx \binstitute  \undefined \def \binstitute#1{#1}\fi
\ifx \binstitutionaled  \undefined \def \binstitutionaled#1{#1}\fi
\ifx \bctitle  \undefined \def \bctitle#1{#1}\fi
\ifx \beditor  \undefined \def \beditor#1{#1}\fi
\ifx \bpublisher  \undefined \def \bpublisher#1{#1}\fi
\ifx \bbtitle  \undefined \def \bbtitle#1{#1}\fi
\ifx \bedition  \undefined \def \bedition#1{#1}\fi
\ifx \bseriesno  \undefined \def \bseriesno#1{#1}\fi
\ifx \blocation  \undefined \def \blocation#1{#1}\fi
\ifx \bsertitle  \undefined \def \bsertitle#1{#1}\fi
\ifx \bsnm \undefined \def \bsnm#1{#1}\fi
\ifx \bsuffix \undefined \def \bsuffix#1{#1}\fi
\ifx \bparticle \undefined \def \bparticle#1{#1}\fi
\ifx \barticle \undefined \def \barticle#1{#1}\fi
\bibcommenthead
\ifx \bconfdate \undefined \def \bconfdate #1{#1}\fi
\ifx \botherref \undefined \def \botherref #1{#1}\fi
\ifx \url \undefined \def \url#1{\textsf{#1}}\fi
\ifx \bchapter \undefined \def \bchapter#1{#1}\fi
\ifx \bbook \undefined \def \bbook#1{#1}\fi
\ifx \bcomment \undefined \def \bcomment#1{#1}\fi
\ifx \oauthor \undefined \def \oauthor#1{#1}\fi
\ifx \citeauthoryear \undefined \def \citeauthoryear#1{#1}\fi
\ifx \endbibitem  \undefined \def \endbibitem {}\fi
\ifx \bconflocation  \undefined \def \bconflocation#1{#1}\fi
\ifx \arxivurl  \undefined \def \arxivurl#1{\textsf{#1}}\fi
\csname PreBibitemsHook\endcsname

\bibitem{dinov2016volume}
\begin{barticle}
\bauthor{\bsnm{Dinov}, \binits{I.D.}}:
\batitle{Volume and value of big healthcare data}.
\bjtitle{Journal of medical statistics and informatics}
\bvolume{4},
\bfpage{3}
(\byear{2016})
\end{barticle}
\endbibitem

\bibitem{galic2008image}
\begin{barticle}
\bauthor{\bsnm{Gali{\'c}}, \binits{I.}},
\bauthor{\bsnm{Weickert}, \binits{J.}},
\bauthor{\bsnm{Welk}, \binits{M.}},
\bauthor{\bsnm{Bruhn}, \binits{A.}},
\bauthor{\bsnm{Belyaev}, \binits{A.}},
\bauthor{\bsnm{Seidel}, \binits{H.-P.}}:
\batitle{Image compression with anisotropic diffusion}.
\bjtitle{Journal of Mathematical Imaging and Vision}
\bvolume{31}(\bissue{2-3}),
\bfpage{255}--\blpage{269}
(\byear{2008})
\end{barticle}
\endbibitem

\bibitem{schmaltz2014understanding}
\begin{barticle}
\bauthor{\bsnm{Schmaltz}, \binits{C.}},
\bauthor{\bsnm{Peter}, \binits{P.}},
\bauthor{\bsnm{Mainberger}, \binits{M.}},
\bauthor{\bsnm{Ebel}, \binits{F.}},
\bauthor{\bsnm{Weickert}, \binits{J.}},
\bauthor{\bsnm{Bruhn}, \binits{A.}}:
\batitle{Understanding, optimising, and extending data compression with
  anisotropic diffusion}.
\bjtitle{Int'l Journal of Computer Vision}
\bvolume{108}(\bissue{3}),
\bfpage{222}--\blpage{240}
(\byear{2014})
\end{barticle}
\endbibitem

\bibitem{peter2015beyond}
\begin{barticle}
\bauthor{\bsnm{Peter}, \binits{P.}},
\bauthor{\bsnm{Schmaltz}, \binits{C.}},
\bauthor{\bsnm{Mach}, \binits{N.}},
\bauthor{\bsnm{Mainberger}, \binits{M.}},
\bauthor{\bsnm{Weickert}, \binits{J.}}:
\batitle{Beyond pure quality: Progressive modes, region of interest coding, and
  real time video decoding for {PDE}-based image compression}.
\bjtitle{Journal of Visual Communication and Image Representation}
\bvolume{31},
\bfpage{253}--\blpage{265}
(\byear{2015})
\end{barticle}
\endbibitem

\bibitem{kostler2007pde}
\begin{botherref}
\oauthor{\bsnm{K{\"o}stler}, \binits{H.}},
\oauthor{\bsnm{St{\"u}rmer}, \binits{M.}},
\oauthor{\bsnm{Freundl}, \binits{C.}},
\oauthor{\bsnm{R{\"u}de}, \binits{U.}}:
{PDE} based video compression in real time.
Technical Report 07-11,
University Erlangen--N{\"u}rnberg, Lehrstuhl f{\"u}r Informatik 10
(2007)
\end{botherref}
\endbibitem

\bibitem{andris2016proof}
\begin{bchapter}
\bauthor{\bsnm{Andris}, \binits{S.}},
\bauthor{\bsnm{Peter}, \binits{P.}},
\bauthor{\bsnm{Weickert}, \binits{J.}}:
\bctitle{A proof-of-concept framework for {PDE}-based video compression}.
In: \bbtitle{Proc.\ Picture Coding Symposium (PCS)},
pp. \bfpage{1}--\blpage{5}
(\byear{2016}).
\bcomment{IEEE}
\end{bchapter}
\endbibitem

\bibitem{10.5555/573326}
\begin{bbook}
\bauthor{\bsnm{Pennebaker}, \binits{W.B.}},
\bauthor{\bsnm{Mitchell}, \binits{J.L.}}:
\bbtitle{JPEG Still Image Data Compression Standard},
\bedition{1st} edn.
\bpublisher{Kluwer Academic Publishers},
\blocation{USA}
(\byear{1992})
\end{bbook}
\endbibitem

\bibitem{Taubman:2002}
\begin{bbook}
\bauthor{\bsnm{Taubman}, \binits{D.}},
\bauthor{\bsnm{Marcellin}, \binits{M.}}:
\bbtitle{JPEG2000: Image Compression Fundamentals, Standards and Practice},
\bedition{1st} edn.
\bpublisher{Springer},
\blocation{USA}
(\byear{2002})
\end{bbook}
\endbibitem

\bibitem{peter2013three}
\begin{bchapter}
\bauthor{\bsnm{Peter}, \binits{P.}}:
\bctitle{Three-dimensional data compression with anisotropic diffusion}.
In: \bbtitle{German Conf.\ on Pattern Recognition},
pp. \bfpage{231}--\blpage{236}
(\byear{2013}).
\bcomment{Springer}
\end{bchapter}
\endbibitem

\bibitem{kil2006lossless}
\begin{barticle}
\bauthor{\bsnm{Kil}, \binits{S.-K.}},
\bauthor{\bsnm{Lee}, \binits{J.-S.}},
\bauthor{\bsnm{Shen}, \binits{D.}},
\bauthor{\bsnm{Ryu}, \binits{J.}},
\bauthor{\bsnm{Lee}, \binits{E.}},
\bauthor{\bsnm{Min}, \binits{H.}},
\bauthor{\bsnm{Hong}, \binits{S.}}:
\batitle{Lossless medical image compression using redundancy analysis}.
\bjtitle{International Journal of Computer Science and Network Security}
\bvolume{6}(\bissue{1}),
\bfpage{50}--\blpage{56}
(\byear{2006})
\end{barticle}
\endbibitem

\bibitem{miaou2009lossless}
\begin{barticle}
\bauthor{\bsnm{Miaou}, \binits{S.-G.}},
\bauthor{\bsnm{Ke}, \binits{F.-S.}},
\bauthor{\bsnm{Chen}, \binits{S.-C.}}:
\batitle{A lossless compression method for medical image sequences using
  {JPEG-LS} and interframe coding}.
\bjtitle{IEEE Transactions on Information Technology in Biomedicine}
\bvolume{13}(\bissue{5}),
\bfpage{818}--\blpage{821}
(\byear{2009})
\end{barticle}
\endbibitem

\bibitem{jumakulyyev2021lossless}
\begin{bchapter}
\bauthor{\bsnm{Jumakulyyev}, \binits{I.}},
\bauthor{\bsnm{Schultz}, \binits{T.}}:
\bctitle{Lossless {PDE}-based compression of {3D} medical images}.
In: \bbtitle{International Conference on Scale Space and Variational Methods in
  Computer Vision},
pp. \bfpage{450}--\blpage{462}
(\byear{2021}).
\bcomment{Springer}
\end{bchapter}
\endbibitem

\bibitem{LeBihan:1986}
\begin{barticle}
\bauthor{\bsnm{Le~Bihan}, \binits{D.}},
\bauthor{\bsnm{Breton}, \binits{E.}},
\bauthor{\bsnm{Lallemand}, \binits{D.}},
\bauthor{\bsnm{Grenier}, \binits{P.}},
\bauthor{\bsnm{Cabanis}, \binits{E.}},
\bauthor{\bsnm{Laval-Jeantet}, \binits{M.}}:
\batitle{{MR} imaging of intravoxel incoherent motions: Application to
  diffusion and perfusion in neurologic disorders}.
\bjtitle{Radiology}
\bvolume{161}(\bissue{2}),
\bfpage{401}--\blpage{407}
(\byear{1986})
\end{barticle}
\endbibitem

\bibitem{Basser:1994}
\begin{barticle}
\bauthor{\bsnm{Basser}, \binits{P.J.}},
\bauthor{\bsnm{Mattiello}, \binits{J.}},
\bauthor{\bsnm{Le~Bihan}, \binits{D.}}:
\batitle{Estimation of the effective self-diffusion tensor from the {NMR} spin
  echo}.
\bjtitle{J.\ of Magnetic Resonance}
\bvolume{B}(\bissue{103}),
\bfpage{247}--\blpage{254}
(\byear{1994})
\end{barticle}
\endbibitem

\bibitem{Jones:2011Book}
\begin{bbook}
\beditor{\bsnm{Jones}, \binits{D.K.}} (ed.):
\bbtitle{Diffusion {MRI}: Theory, Method, and Applications},
\bedition{1st} edn.
\bpublisher{Oxford University Press},
\blocation{United Kingdom}
(\byear{2011})
\end{bbook}
\endbibitem

\bibitem{10.17487/RFC1952}
\begin{botherref}
\oauthor{\bsnm{Deutsch}, \binits{P.}}:
{RFC1952}: {GZIP} File Format Specification Version 4.3.
RFC Editor,
USA
(1996)
\end{botherref}
\endbibitem

\bibitem{iijima1962basic}
\begin{barticle}
\bauthor{\bsnm{Iijima}, \binits{T.}}:
\batitle{Basic theory on the normalization of pattern (in case of typical
  one-dimensional pattern)}.
\bjtitle{Bulletin of Electro-technical Laboratory}
\bvolume{26},
\bfpage{368}--\blpage{388}
(\byear{1962})
\end{barticle}
\endbibitem

\bibitem{perona1990scale}
\begin{barticle}
\bauthor{\bsnm{Perona}, \binits{P.}},
\bauthor{\bsnm{Malik}, \binits{J.}}:
\batitle{Scale-space and edge detection using anisotropic diffusion}.
\bjtitle{IEEE Trans.\ on Pattern Analysis and Machine Intelligence}
\bvolume{12}(\bissue{7}),
\bfpage{629}--\blpage{639}
(\byear{1990})
\end{barticle}
\endbibitem

\bibitem{osher1990feature}
\begin{barticle}
\bauthor{\bsnm{Osher}, \binits{S.}},
\bauthor{\bsnm{Rudin}, \binits{L.I.}}:
\batitle{Feature-oriented image enhancement using shock filters}.
\bjtitle{SIAM Journal on numerical analysis}
\bvolume{27}(\bissue{4}),
\bfpage{919}--\blpage{940}
(\byear{1990})
\end{barticle}
\endbibitem

\bibitem{weickert1998anisotropic}
\begin{bbook}
\bauthor{\bsnm{Weickert}, \binits{J.}}:
\bbtitle{Anisotropic Diffusion in Image Processing}.
\bpublisher{Teubner},
\blocation{Stuttgart}
(\byear{1998})
\end{bbook}
\endbibitem

\bibitem{masnou1998level}
\begin{bchapter}
\bauthor{\bsnm{Masnou}, \binits{S.}},
\bauthor{\bsnm{Morel}, \binits{J.-M.}}:
\bctitle{Level lines based disocclusion}.
In: \bbtitle{Proc.\ Int'l Conf.\ on Image Processing (ICIP)},
pp. \bfpage{259}--\blpage{263}
(\byear{1998}).
\bcomment{IEEE}
\end{bchapter}
\endbibitem

\bibitem{bertalmio2000image}
\begin{bchapter}
\bauthor{\bsnm{Bertalmio}, \binits{M.}},
\bauthor{\bsnm{Sapiro}, \binits{G.}},
\bauthor{\bsnm{Caselles}, \binits{V.}},
\bauthor{\bsnm{Ballester}, \binits{C.}}:
\bctitle{Image inpainting}.
In: \bbtitle{Proc.\ Conf.\ on Computer Graphics and Interactive Techniques
  (SIGGRAPH)},
pp. \bfpage{417}--\blpage{424}
(\byear{2000})
\end{bchapter}
\endbibitem

\bibitem{chan2001nontexture}
\begin{barticle}
\bauthor{\bsnm{Chan}, \binits{T.F.}},
\bauthor{\bsnm{Shen}, \binits{J.}}:
\batitle{Nontexture inpainting by curvature-driven diffusions}.
\bjtitle{Journal of Visual Communication and Image Representation}
\bvolume{12}(\bissue{4}),
\bfpage{436}--\blpage{449}
(\byear{2001})
\end{barticle}
\endbibitem

\bibitem{mainberger2011edge}
\begin{barticle}
\bauthor{\bsnm{Mainberger}, \binits{M.}},
\bauthor{\bsnm{Bruhn}, \binits{A.}},
\bauthor{\bsnm{Weickert}, \binits{J.}},
\bauthor{\bsnm{Forchhammer}, \binits{S.}}:
\batitle{Edge-based compression of cartoon-like images with homogeneous
  diffusion}.
\bjtitle{Pattern Recognition}
\bvolume{44}(\bissue{9}),
\bfpage{1859}--\blpage{1873}
(\byear{2011})
\end{barticle}
\endbibitem

\bibitem{jost2020compressing}
\begin{bchapter}
\bauthor{\bsnm{Jost}, \binits{F.}},
\bauthor{\bsnm{Peter}, \binits{P.}},
\bauthor{\bsnm{Weickert}, \binits{J.}}:
\bctitle{Compressing flow fields with edge-aware homogeneous diffusion
  inpainting}.
In: \bbtitle{IEEE Int'l Conf.\ on Acoustics, Speech and Signal Processing
  (ICASSP)},
pp. \bfpage{2198}--\blpage{2202}
(\byear{2020})
\end{bchapter}
\endbibitem

\bibitem{gautier2012efficient}
\begin{bchapter}
\bauthor{\bsnm{Gautier}, \binits{J.}},
\bauthor{\bsnm{Le~Meur}, \binits{O.}},
\bauthor{\bsnm{Guillemot}, \binits{C.}}:
\bctitle{Efficient depth map compression based on lossless edge coding and
  diffusion}.
In: \bbtitle{2012 Picture Coding Symposium},
pp. \bfpage{81}--\blpage{84}
(\byear{2012}).
\bcomment{IEEE}
\end{bchapter}
\endbibitem

\bibitem{hoffmann2013compression}
\begin{bchapter}
\bauthor{\bsnm{Hoffmann}, \binits{S.}},
\bauthor{\bsnm{Mainberger}, \binits{M.}},
\bauthor{\bsnm{Weickert}, \binits{J.}},
\bauthor{\bsnm{Puhl}, \binits{M.}}:
\bctitle{Compression of depth maps with segment-based homogeneous diffusion}.
In: \bbtitle{Int'l Conf.\ on Scale Space and Variational Methods in Computer
  Vision},
pp. \bfpage{319}--\blpage{330}
(\byear{2013}).
\bcomment{Springer}
\end{bchapter}
\endbibitem

\bibitem{li2012scalable}
\begin{bchapter}
\bauthor{\bsnm{Li}, \binits{Y.}},
\bauthor{\bsnm{Sj{\"o}str{\"o}m}, \binits{M.}},
\bauthor{\bsnm{Jennehag}, \binits{U.}},
\bauthor{\bsnm{Olsson}, \binits{R.}}:
\bctitle{A scalable coding approach for high quality depth image compression}.
In: \bbtitle{Proc.\ 3DTV-Conference: The True Vision-Capture, Transmission and
  Display of 3D Video (3DTV-CON)},
pp. \bfpage{1}--\blpage{4}
(\byear{2012}).
\bcomment{IEEE}
\end{bchapter}
\endbibitem

\bibitem{charbonnier1997deterministic}
\begin{barticle}
\bauthor{\bsnm{{Charbonnier}}, \binits{P.}},
\bauthor{\bsnm{{Blanc{-}F{\'{e}}raud}}, \binits{L.}},
\bauthor{\bsnm{{Aubert}}, \binits{G.}},
\bauthor{\bsnm{{Barlaud}}, \binits{M.}}:
\batitle{Deterministic edge-preserving regularization in computed imaging}.
\bjtitle{IEEE Trans.\ on Image Processing}
\bvolume{6}(\bissue{2}),
\bfpage{298}--\blpage{311}
(\byear{1997})
\end{barticle}
\endbibitem

\bibitem{10.1007/978-3-7091-6586-7_13}
\begin{bchapter}
\bauthor{\bsnm{Weickert}, \binits{J.}}:
\bctitle{Theoretical foundations of anisotropic diffusion in image processing}.
In: \beditor{\bsnm{Kropatsch}, \binits{W.}},
\beditor{\bsnm{Klette}, \binits{R.}},
\beditor{\bsnm{Solina}, \binits{F.}},
\beditor{\bsnm{Albrecht}, \binits{R.}} (eds.)
\bbtitle{Theoretical Foundations of Computer Vision},
pp. \bfpage{221}--\blpage{236}.
\bpublisher{Springer},
\blocation{Vienna}
(\byear{1996})
\end{bchapter}
\endbibitem

\bibitem{you2000fourth}
\begin{barticle}
\bauthor{\bsnm{You}, \binits{Y.-L.}},
\bauthor{\bsnm{Kaveh}, \binits{M.}}:
\batitle{Fourth-order partial differential equations for noise removal}.
\bjtitle{IEEE Trans.\ on Image Processing}
\bvolume{9}(\bissue{10}),
\bfpage{1723}--\blpage{1730}
(\byear{2000})
\end{barticle}
\endbibitem

\bibitem{lysaker2003noise}
\begin{barticle}
\bauthor{\bsnm{{Lysaker}}, \binits{M.}},
\bauthor{\bsnm{{Lundervold}}, \binits{A.}},
\bauthor{\bsnm{{Tai}}, \binits{X.-C.}}:
\batitle{Noise removal using fourth-order partial differential equation with
  applications to medical magnetic resonance images in space and time}.
\bjtitle{IEEE Trans.\ on Image Processing}
\bvolume{12}(\bissue{12}),
\bfpage{1579}--\blpage{1590}
(\byear{2003})
\end{barticle}
\endbibitem

\bibitem{didas2009properties}
\begin{barticle}
\bauthor{\bsnm{{Didas}}, \binits{S.}},
\bauthor{\bsnm{{Weickert}}, \binits{J.}},
\bauthor{\bsnm{{Burgeth}}, \binits{B.}}:
\batitle{Properties of higher order nonlinear diffusion filtering}.
\bjtitle{Journal of Mathematical Imaging and Vision}
\bvolume{35}(\bissue{3}),
\bfpage{208}--\blpage{226}
(\byear{2009})
\end{barticle}
\endbibitem

\bibitem{hajiaboli2011anisotropic}
\begin{barticle}
\bauthor{\bsnm{Hajiaboli}, \binits{M.R.}}:
\batitle{An anisotropic fourth-order diffusion filter for image noise removal}.
\bjtitle{Int'l Journal of Computer Vision}
\bvolume{92}(\bissue{2}),
\bfpage{177}--\blpage{191}
(\byear{2011})
\end{barticle}
\endbibitem

\bibitem{zadeh2017multi}
\begin{barticle}
\bauthor{\bsnm{Gorgi~Zadeh}, \binits{S.}},
\bauthor{\bsnm{{Didas}}, \binits{S.}},
\bauthor{\bsnm{{Wintergerst}}, \binits{M.W.M.}},
\bauthor{\bsnm{{Schultz}}, \binits{T.}}:
\batitle{Multi-scale anisotropic fourth-order diffusion improves ridge and
  valley localization}.
\bjtitle{Journal of Mathematical Imaging and Vision}
\bvolume{59}(\bissue{2}),
\bfpage{257}--\blpage{269}
(\byear{2017})
\end{barticle}
\endbibitem

\bibitem{li2013two}
\begin{barticle}
\bauthor{\bsnm{{Li}}, \binits{P.}},
\bauthor{\bsnm{{Li}}, \binits{S.-J.}},
\bauthor{\bsnm{{Yao}}, \binits{Z.-A.}},
\bauthor{\bsnm{{Zhang}}, \binits{Z.-J.}}:
\batitle{Two anisotropic fourth-order partial differential equations for image
  inpainting}.
\bjtitle{IET Image Processing}
\bvolume{7}(\bissue{3}),
\bfpage{260}--\blpage{269}
(\byear{2013})
\end{barticle}
\endbibitem

\bibitem{10.1007/978-3-030-56215-1_5}
\begin{bchapter}
\bauthor{\bsnm{Jumakulyyev}, \binits{I.}},
\bauthor{\bsnm{Schultz}, \binits{T.}}:
\bctitle{Fourth-order anisotropic diffusion for inpainting and image
  compression}.
In: \beditor{\bsnm{{\"O}zarslan}, \binits{E.}},
\beditor{\bsnm{Schultz}, \binits{T.}},
\beditor{\bsnm{Zhang}, \binits{E.}},
\beditor{\bsnm{Fuster}, \binits{A.}} (eds.)
\bbtitle{Anisotropy Across Fields and Scales},
pp. \bfpage{99}--\blpage{124}.
\bpublisher{Springer},
\blocation{Cham}
(\byear{2021})
\end{bchapter}
\endbibitem

\bibitem{chen2014bi}
\begin{bchapter}
\bauthor{\bsnm{Chen}, \binits{Y.}},
\bauthor{\bsnm{Ranftl}, \binits{R.}},
\bauthor{\bsnm{Pock}, \binits{T.}}:
\bctitle{A bi-level view of inpainting-based image compression}.
In: \beditor{\bsnm{K{\'u}kelov{\'a}}, \binits{Z.}},
\beditor{\bsnm{Heller}, \binits{J.}} (eds.)
\bbtitle{Proc.\ Computer Vision Winter Workshop},
pp. \bfpage{19}--\blpage{26}
(\byear{2014})
\end{bchapter}
\endbibitem

\bibitem{peter2016evaluating}
\begin{barticle}
\bauthor{\bsnm{Peter}, \binits{P.}},
\bauthor{\bsnm{Hoffmann}, \binits{S.}},
\bauthor{\bsnm{Nedwed}, \binits{F.}},
\bauthor{\bsnm{Hoeltgen}, \binits{L.}},
\bauthor{\bsnm{Weickert}, \binits{J.}}:
\batitle{Evaluating the true potential of diffusion-based inpainting in a
  compression context}.
\bjtitle{Signal Processing: Image Communication}
\bvolume{46},
\bfpage{40}--\blpage{53}
(\byear{2016})
\end{barticle}
\endbibitem

\bibitem{amrani2017diffusion}
\begin{barticle}
\bauthor{\bsnm{Amrani}, \binits{N.}},
\bauthor{\bsnm{Serra-Sagrist{\`a}}, \binits{J.}},
\bauthor{\bsnm{Peter}, \binits{P.}},
\bauthor{\bsnm{Weickert}, \binits{J.}}:
\batitle{Diffusion-based inpainting for coding remote-sensing data}.
\bjtitle{IEEE Geoscience and Remote Sensing Letters}
\bvolume{14}(\bissue{8}),
\bfpage{1203}--\blpage{1207}
(\byear{2017})
\end{barticle}
\endbibitem

\bibitem{peter2016turning}
\begin{barticle}
\bauthor{\bsnm{Peter}, \binits{P.}},
\bauthor{\bsnm{Kaufhold}, \binits{L.}},
\bauthor{\bsnm{Weickert}, \binits{J.}}:
\batitle{Turning diffusion-based image colorization into efficient color
  compression}.
\bjtitle{IEEE Transactions on Image Processing}
\bvolume{26}(\bissue{2}),
\bfpage{860}--\blpage{869}
(\byear{2016})
\end{barticle}
\endbibitem

\bibitem{Beaulieu:2002}
\begin{barticle}
\bauthor{\bsnm{Beaulieu}, \binits{C.}}:
\batitle{The basis of anisotropic water diffusion in the nervous system -- a
  technical review}.
\bjtitle{NMR in Biomedicine}
\bvolume{15}(\bissue{7--8}),
\bfpage{435}--\blpage{455}
(\byear{2002})
\end{barticle}
\endbibitem

\bibitem{stejskal1965spin}
\begin{barticle}
\bauthor{\bsnm{Stejskal}, \binits{E.O.}},
\bauthor{\bsnm{Tanner}, \binits{J.E.}}:
\batitle{Spin diffusion measurements: spin echoes in the presence of a
  time-dependent field gradient}.
\bjtitle{The journal of chemical physics}
\bvolume{42}(\bissue{1}),
\bfpage{288}--\blpage{292}
(\byear{1965})
\end{barticle}
\endbibitem

\bibitem{Pierpaoli:1996a}
\begin{barticle}
\bauthor{\bsnm{Pierpaoli}, \binits{C.}},
\bauthor{\bsnm{Basser}, \binits{P.J.}}:
\batitle{Toward a quantitative assessment of diffusion anisotropy}.
\bjtitle{Magnetic Resonance in Medicine}
\bvolume{36},
\bfpage{893}--\blpage{906}
(\byear{1996})
\end{barticle}
\endbibitem

\bibitem{Callaghan:1988}
\begin{barticle}
\bauthor{\bsnm{Callaghan}, \binits{P.T.}},
\bauthor{\bsnm{Eccles}, \binits{C.D.}},
\bauthor{\bsnm{Xia}, \binits{Y.}}:
\batitle{{NMR} microscopy of dynamic displacements: k-space and q-space
  imaging}.
\bjtitle{Journal of Physics E}
\bvolume{21}(\bissue{8}),
\bfpage{820}--\blpage{822}
(\byear{1988})
\end{barticle}
\endbibitem

\bibitem{cheng2017single}
\begin{barticle}
\bauthor{\bsnm{Cheng}, \binits{J.}},
\bauthor{\bsnm{Shen}, \binits{D.}},
\bauthor{\bsnm{Yap}, \binits{P.-T.}},
\bauthor{\bsnm{Basser}, \binits{P.J.}}:
\batitle{Single-and multiple-shell uniform sampling schemes for diffusion {MRI}
  using spherical codes}.
\bjtitle{IEEE Transactions on Medical Imaging}
\bvolume{37}(\bissue{1}),
\bfpage{185}--\blpage{199}
(\byear{2017})
\end{barticle}
\endbibitem

\bibitem{Andersson2016a}
\begin{barticle}
\bauthor{\bsnm{Andersson}, \binits{J.L.R.}},
\bauthor{\bsnm{Sotiropoulos}, \binits{S.N.}}:
\batitle{An integrated approach to correction for off-resonance effects and
  subject movement in diffusion {MR} imaging}.
\bjtitle{{NeuroImage}}
\bvolume{125},
\bfpage{1063}--\blpage{1078}
(\byear{2016})
\end{barticle}
\endbibitem

\bibitem{kassim2005motion}
\begin{barticle}
\bauthor{\bsnm{Kassim}, \binits{A.A.}},
\bauthor{\bsnm{Yan}, \binits{P.}},
\bauthor{\bsnm{Lee}, \binits{W.S.}},
\bauthor{\bsnm{Sengupta}, \binits{K.}}:
\batitle{Motion compensated lossy-to-lossless compression of {4-D} medical
  images using integer wavelet transforms}.
\bjtitle{IEEE Trans.\ on Information Technology in Biomedicine}
\bvolume{9}(\bissue{1}),
\bfpage{132}--\blpage{138}
(\byear{2005})
\end{barticle}
\endbibitem

\bibitem{sanchez2008efficient}
\begin{barticle}
\bauthor{\bsnm{Sanchez}, \binits{V.}},
\bauthor{\bsnm{Nasiopoulos}, \binits{P.}},
\bauthor{\bsnm{Abugharbieh}, \binits{R.}}:
\batitle{Efficient lossless compression of {4-D} medical images based on the
  advanced video coding scheme}.
\bjtitle{IEEE Trans.\ on Information Technology in Biomedicine}
\bvolume{12}(\bissue{4}),
\bfpage{442}--\blpage{446}
(\byear{2008})
\end{barticle}
\endbibitem

\bibitem{zeng2002four}
\begin{barticle}
\bauthor{\bsnm{Zeng}, \binits{L.}},
\bauthor{\bsnm{Jansen}, \binits{C.P.}},
\bauthor{\bsnm{Marsch}, \binits{S.}},
\bauthor{\bsnm{Unser}, \binits{M.}},
\bauthor{\bsnm{Hunziker}, \binits{P.R.}}:
\batitle{Four-dimensional wavelet compression of arbitrarily sized
  echocardiographic data}.
\bjtitle{IEEE Transactions on Medical Imaging}
\bvolume{21}(\bissue{9}),
\bfpage{1179}--\blpage{1187}
(\byear{2002})
\end{barticle}
\endbibitem

\bibitem{lalgudi2005compression}
\begin{bchapter}
\bauthor{\bsnm{Lalgudi}, \binits{H.G.}},
\bauthor{\bsnm{Bilgin}, \binits{A.}},
\bauthor{\bsnm{Marcellin}, \binits{M.W.}},
\bauthor{\bsnm{Nadar}, \binits{M.S.}}:
\bctitle{Compression of {fMRI} and ultrasound images using {4D} {SPIHT}}.
In: \bbtitle{IEEE Int'l Conf.\ on Image Processing (ICIP)},
vol. \bseriesno{2},
pp. \bfpage{746}--\blpage{749}
(\byear{2005})
\end{bchapter}
\endbibitem

\bibitem{liu2007four}
\begin{bchapter}
\bauthor{\bsnm{Liu}, \binits{Y.}},
\bauthor{\bsnm{Pearlman}, \binits{W.A.}}:
\bctitle{Four-dimensional wavelet compression of {4-D} medical images using
  scalable {4-D} {SBHP}}.
In: \bbtitle{Proc.\ Data Compression Conference (DCC)},
pp. \bfpage{233}--\blpage{242}
(\byear{2007}).
\bcomment{IEEE}
\end{bchapter}
\endbibitem

\bibitem{belhadef2016lossless}
\begin{barticle}
\bauthor{\bsnm{Belhadef}, \binits{L.}},
\bauthor{\bsnm{Maaza}, \binits{Z.M.}}:
\batitle{Lossless {4D} medical images compression with motion compensation and
  lifting wavelet transform}.
\bjtitle{Int.\ J.\ Signal Process.\ Syst}
\bvolume{4}(\bissue{2}),
\bfpage{168}--\blpage{171}
(\byear{2016})
\end{barticle}
\endbibitem

\bibitem{nguyen2011efficient}
\begin{barticle}
\bauthor{\bsnm{Nguyen}, \binits{B.P.}},
\bauthor{\bsnm{Chui}, \binits{C.-K.}},
\bauthor{\bsnm{Ong}, \binits{S.-H.}},
\bauthor{\bsnm{Chang}, \binits{S.}}:
\batitle{An efficient compression scheme for {4-D} medical images using
  hierarchical vector quantization and motion compensation}.
\bjtitle{Computers in Biology and Medicine}
\bvolume{41}(\bissue{9}),
\bfpage{843}--\blpage{856}
(\byear{2011})
\end{barticle}
\endbibitem

\bibitem{Witten:1987}
\begin{barticle}
\bauthor{\bsnm{Witten}, \binits{I.H.}},
\bauthor{\bsnm{Neal}, \binits{R.M.}},
\bauthor{\bsnm{Cleary}, \binits{J.G.}}:
\batitle{Arithmetic coding for data compression}.
\bjtitle{Communications of the {ACM}}
\bvolume{30}(\bissue{6}),
\bfpage{520}--\blpage{540}
(\byear{1987})
\end{barticle}
\endbibitem

\bibitem{Duda:2015}
\begin{bchapter}
\bauthor{\bsnm{Duda}, \binits{J.}},
\bauthor{\bsnm{Tahboub}, \binits{K.}},
\bauthor{\bsnm{Gadgil}, \binits{N.J.}},
\bauthor{\bsnm{Delp}, \binits{E.J.}}:
\bctitle{The use of asymmetric numeral systems as an accurate replacement for
  {Huffman} coding}.
In: \bbtitle{Proc.\ IEEE Picture Coding Symposium {(PCS)}},
pp. \bfpage{65}--\blpage{69}
(\byear{2015})
\end{bchapter}
\endbibitem

\bibitem{Merlet:2013}
\begin{botherref}
\oauthor{\bsnm{Merlet}, \binits{S.}}:
Compressive sensing in diffusion {MRI}.
PhD thesis,
Université Nice Sophia Antipolis
(2013)
\end{botherref}
\endbibitem

\bibitem{Tobisch:2018Frontiers}
\begin{barticle}
\bauthor{\bsnm{Tobisch}, \binits{A.}},
\bauthor{\bsnm{Stirnberg}, \binits{R.}},
\bauthor{\bsnm{Harms}, \binits{R.L.}},
\bauthor{\bsnm{Schultz}, \binits{T.}},
\bauthor{\bsnm{Roebroeck}, \binits{A.}},
\bauthor{\bsnm{Breteler}, \binits{M.M.}},
\bauthor{\bsnm{St{\"o}cker}, \binits{T.}}:
\batitle{Compressed sensing diffusion spectrum imaging for accelerated
  diffusion microstructure {MRI} in long-term population imaging}.
\bjtitle{Frontiers in Neuroscience}
\bvolume{12},
\bfpage{650}
(\byear{2018})
\end{barticle}
\endbibitem

\bibitem{nagoor2020lossless}
\begin{bchapter}
\bauthor{\bsnm{Nagoor}, \binits{O.H.}},
\bauthor{\bsnm{Whittle}, \binits{J.}},
\bauthor{\bsnm{Deng}, \binits{J.}},
\bauthor{\bsnm{Mora}, \binits{B.}},
\bauthor{\bsnm{Jones}, \binits{M.W.}}:
\bctitle{Lossless compression for volumetric medical images using deep neural
  network with local sampling}.
In: \bbtitle{Proc.\ IEEE Int'l Conf.\ on Image Processing (ICIP)},
pp. \bfpage{2815}--\blpage{2819}
(\byear{2020})
\end{bchapter}
\endbibitem

\bibitem{Nasr:2019}
\begin{bchapter}
\bauthor{\bsnm{Nasr}, \binits{M.}},
\bauthor{\bsnm{Shokri}, \binits{R.}},
\bauthor{\bsnm{Houmansadr}, \binits{A.}}:
\bctitle{Comprehensive privacy analysis of deep learning: Passive and active
  white-box inference attacks against centralized and federated learning}.
In: \bbtitle{Proc.\ {IEEE} Symp.\ on Security and Privacy},
pp. \bfpage{739}--\blpage{753}
(\byear{2019})
\end{bchapter}
\endbibitem

\bibitem{marwood2018representing}
\begin{bchapter}
\bauthor{\bsnm{Marwood}, \binits{D.}},
\bauthor{\bsnm{Massimino}, \binits{P.}},
\bauthor{\bsnm{Covell}, \binits{M.}},
\bauthor{\bsnm{Baluja}, \binits{S.}}:
\bctitle{Representing images in 200 bytes: Compression via triangulation}.
In: \bbtitle{2018 25th IEEE International Conference on Image Processing
  (ICIP)},
pp. \bfpage{405}--\blpage{409}
(\byear{2018}).
\bcomment{IEEE}
\end{bchapter}
\endbibitem

\bibitem{peter2019fast}
\begin{bchapter}
\bauthor{\bsnm{Peter}, \binits{P.}}:
\bctitle{Fast inpainting-based compression: Combining shepard interpolation
  with joint inpainting and prediction}.
In: \bbtitle{2019 IEEE International Conference on Image Processing (ICIP)},
pp. \bfpage{3557}--\blpage{3561}
(\byear{2019}).
\bcomment{IEEE}
\end{bchapter}
\endbibitem

\bibitem{Chizhov:2021}
\begin{bchapter}
\bauthor{\bsnm{Chizhov}, \binits{V.}},
\bauthor{\bsnm{Weickert}, \binits{J.}}:
\bctitle{Efficient data optimisation for harmonic inpainting with finite
  elements}.
In: \bbtitle{Int'l Conf.\ on Computer Analysis of Images and Patterns},
pp. \bfpage{432}--\blpage{441}
(\byear{2021})
\end{bchapter}
\endbibitem

\bibitem{logg2012automated}
\begin{bbook}
\bauthor{\bsnm{Logg}, \binits{A.}},
\bauthor{\bsnm{Mardal}, \binits{K.-A.}},
\bauthor{\bsnm{Wells}, \binits{G.}}:
\bbtitle{Automated Solution of Differential Equations by the Finite Element
  Method: The {FEniCS} Book}.
\bsertitle{Lecture Notes in Computational Science and Engineering},
vol. \bseriesno{84}.
\bpublisher{Springer},
\blocation{Berlin, Heidelberg}
(\byear{2012})
\end{bbook}
\endbibitem

\bibitem{mainberger2011optimising}
\begin{bchapter}
\bauthor{\bsnm{Mainberger}, \binits{M.}},
\bauthor{\bsnm{Hoffmann}, \binits{S.}},
\bauthor{\bsnm{Weickert}, \binits{J.}},
\bauthor{\bsnm{Tang}, \binits{C.H.}},
\bauthor{\bsnm{Johannsen}, \binits{D.}},
\bauthor{\bsnm{Neumann}, \binits{F.}},
\bauthor{\bsnm{Doerr}, \binits{B.}}:
\bctitle{Optimising spatial and tonal data for homogeneous diffusion
  inpainting}.
In: \bbtitle{International Conference on Scale Space and Variational Methods in
  Computer Vision},
pp. \bfpage{26}--\blpage{37}
(\byear{2011}).
\bcomment{Springer}
\end{bchapter}
\endbibitem

\bibitem{langtangen2017solving}
\begin{bbook}
\bauthor{\bsnm{Langtangen}, \binits{H.P.}},
\bauthor{\bsnm{Logg}, \binits{A.}}:
\bbtitle{Solving PDEs in Python: The FEniCS Tutorial I},
\bedition{1st} edn.
\bpublisher{Springer},
\blocation{Oslo}
(\byear{2017})
\end{bbook}
\endbibitem

\bibitem{2020SciPy-NMeth}
\begin{barticle}
\bauthor{\bsnm{Virtanen}, \binits{P.}},
\bauthor{\bsnm{Gommers}, \binits{R.}},
\bauthor{\bsnm{Oliphant}, \binits{T.E.}},
\bauthor{\bsnm{Haberland}, \binits{M.}},
\bauthor{\bsnm{Reddy}, \binits{T.}},
\bauthor{\bsnm{Cournapeau}, \binits{D.}},
\bauthor{\bsnm{Burovski}, \binits{E.}},
\bauthor{\bsnm{Peterson}, \binits{P.}},
\bauthor{\bsnm{Weckesser}, \binits{W.}},
\bauthor{\bsnm{Bright}, \binits{J.}},
\bauthor{\bsnm{{van der Walt}}, \binits{S.J.}},
\bauthor{\bsnm{Brett}, \binits{M.}},
\bauthor{\bsnm{Wilson}, \binits{J.}},
\bauthor{\bsnm{Millman}, \binits{K.J.}},
\bauthor{\bsnm{Mayorov}, \binits{N.}},
\bauthor{\bsnm{Nelson}, \binits{A.R.J.}},
\bauthor{\bsnm{Jones}, \binits{E.}},
\bauthor{\bsnm{Kern}, \binits{R.}},
\bauthor{\bsnm{Larson}, \binits{E.}},
\bauthor{\bsnm{Carey}, \binits{C.J.}},
\bauthor{\bsnm{Polat}, \binits{{\. I}.}},
\bauthor{\bsnm{Feng}, \binits{Y.}},
\bauthor{\bsnm{Moore}, \binits{E.W.}},
\bauthor{\bsnm{{VanderPlas}}, \binits{J.}},
\bauthor{\bsnm{Laxalde}, \binits{D.}},
\bauthor{\bsnm{Perktold}, \binits{J.}},
\bauthor{\bsnm{Cimrman}, \binits{R.}},
\bauthor{\bsnm{Henriksen}, \binits{I.}},
\bauthor{\bsnm{Quintero}, \binits{E.A.}},
\bauthor{\bsnm{Harris}, \binits{C.R.}},
\bauthor{\bsnm{Archibald}, \binits{A.M.}},
\bauthor{\bsnm{Ribeiro}, \binits{A.H.}},
\bauthor{\bsnm{Pedregosa}, \binits{F.}},
\bauthor{\bsnm{{van Mulbregt}}, \binits{P.}},
\bauthor{\bsnm{{SciPy 1.0 Contributors}}}:
\batitle{{{SciPy} 1.0: Fundamental Algorithms for Scientific Computing in
  Python}}.
\bjtitle{Nature Methods}
\bvolume{17},
\bfpage{261}--\blpage{272}
(\byear{2020})
\end{barticle}
\endbibitem

\bibitem{jenkinson2002improved}
\begin{barticle}
\bauthor{\bsnm{Jenkinson}, \binits{M.}},
\bauthor{\bsnm{Bannister}, \binits{P.}},
\bauthor{\bsnm{Brady}, \binits{M.}},
\bauthor{\bsnm{Smith}, \binits{S.}}:
\batitle{Improved optimization for the robust and accurate linear registration
  and motion correction of brain images}.
\bjtitle{Neuroimage}
\bvolume{17}(\bissue{2}),
\bfpage{825}--\blpage{841}
(\byear{2002})
\end{barticle}
\endbibitem

\bibitem{leemans2009b}
\begin{barticle}
\bauthor{\bsnm{Leemans}, \binits{A.}},
\bauthor{\bsnm{Jones}, \binits{D.K.}}:
\batitle{The {B}-matrix must be rotated when correcting for subject motion in
  {DTI} data}.
\bjtitle{Magnetic Resonance in Medicine}
\bvolume{61}(\bissue{6}),
\bfpage{1336}--\blpage{1349}
(\byear{2009})
\end{barticle}
\endbibitem

\bibitem{garyfallidis2014dipy}
\begin{barticle}
\bauthor{\bsnm{Garyfallidis}, \binits{E.}},
\bauthor{\bsnm{Brett}, \binits{M.}},
\bauthor{\bsnm{Amirbekian}, \binits{B.}},
\bauthor{\bsnm{Rokem}, \binits{A.}},
\bauthor{\bsnm{Van Der~Walt}, \binits{S.}},
\bauthor{\bsnm{Descoteaux}, \binits{M.}},
\bauthor{\bsnm{Nimmo-Smith}, \binits{I.}},
\bauthor{\bsnm{{Dipy Contributors}}}:
\batitle{Dipy, a library for the analysis of diffusion {MRI} data}.
\bjtitle{Frontiers in Neuroinformatics}
\bvolume{8},
\bfpage{8}
(\byear{2014})
\end{barticle}
\endbibitem

\bibitem{koch2019shore}
\begin{barticle}
\bauthor{\bsnm{Koch}, \binits{A.}},
\bauthor{\bsnm{Zhukov}, \binits{A.}},
\bauthor{\bsnm{St{\"o}cker}, \binits{T.}},
\bauthor{\bsnm{Groeschel}, \binits{S.}},
\bauthor{\bsnm{Schultz}, \binits{T.}}:
\batitle{{SHORE}-based detection and imputation of dropout in diffusion {MRI}}.
\bjtitle{Magnetic Resonance in Medicine}
\bvolume{82}(\bissue{6}),
\bfpage{2286}--\blpage{2298}
(\byear{2019})
\end{barticle}
\endbibitem

\bibitem{Alexander:2002}
\begin{barticle}
\bauthor{\bsnm{Alexander}, \binits{D.C.}},
\bauthor{\bsnm{Barker}, \binits{G.J.}},
\bauthor{\bsnm{Arridge}, \binits{S.R.}}:
\batitle{Detection and modeling of non-gaussian apparent diffusion coefficient
  profiles in human brain data}.
\bjtitle{Magnetic Resonance in Medicine}
\bvolume{48},
\bfpage{331}--\blpage{340}
(\byear{2002})
\end{barticle}
\endbibitem

\end{thebibliography}
\end{document}